\documentclass[reprint,10pt,nofootinbib,twocolumn,aps,prx,preprintnumbers,citeautoscript,longbibliography]{revtex4-2}
\usepackage{physics}
\usepackage{makecell}
\usepackage{graphicx}
\usepackage{amsmath}
\usepackage[hidelinks,hypertexnames=false]{hyperref}
\usepackage{dsfont}
\usepackage[ddmmyyyy,hhmmss]{datetime}
\usepackage{tipa}
\usepackage{booktabs}
\usepackage{braket}
\usepackage{xcolor}
\usepackage{lineno}
\usepackage{siunitx}
\usepackage{soul}
\usepackage{verbatim}
\usepackage{array}

\long\def\symbolfootnote[#1]#2{\begingroup%
\def\thefootnote{\fnsymbol{footnote}}\footnotetext[#1]{#2}\endgroup}

\usepackage{titlesec}
\titleformat{\section}
  {\centering\normalfont\normalsize\scshape}{\thesection.}{1em}{}
\titleformat{\subsection}
  {\centering\normalfont\normalsize\bfseries}{\thesubsection.}{1em}{}
\titleformat{\subsubsection}
  {\normalfont\normalsize\bfseries\itshape}{\thesubsubsection.}{1em}{}

\begin{document} 

\title{Probing supersolidity through excitations in a spin-orbit-coupled Bose-Einstein condensate}

\author{C. S. Chisholm$^{1,\ast,\dagger}$}
\author{S. Hirthe$^{1,\ast}$}
\author{V. B. Makhalov$^{1,\ast,\ddagger}$}
\author{R. Ramos$^{1,\ast,\ddagger}$}
\author{R. Vatré$^{1,\ast,\S}$}
\author{J. Cabedo$^{2}$}
\author{A. Celi$^{2}$}
\author{L. Tarruell$^{1,3,\P}$}

\affiliation{$^1$ICFO - Institut de Ciencies Fotoniques, The Barcelona Institute of Science and Technology, 08860 Castelldefels (Barcelona), Spain}
\affiliation{$^2$Departament de F\'{i}sica, Universitat Aut\`{o}noma de Barcelona, E-08193 Bellaterra, Spain}
\affiliation{$^3$ICREA, Pg. Llu\'{i}s Companys 23, 08010 Barcelona, Spain}

\symbolfootnote[1]{These authors contributed equally to this work. They are listed in alphabetic order.}
\symbolfootnote[2]{Present address: OpenStar Technologies, 20 Glover Street, Ngauranga, Wellington, New Zealand.}
\symbolfootnote[3]{Present address: Ideaded, Carrer de la Tecnologia, 35, Viladecans, Barcelona, Spain.}
\symbolfootnote[4]{Present address: Université Sorbonne Paris Nord, Laboratoire de Physique des Lasers, 99 av. J.-B. Cl\'{e}ment, Villetaneuse, France.}
\symbolfootnote[5]{Electronic address: leticia.tarruell@icfo.eu}

\begin{abstract}
Spin-orbit-coupled Bose-Einstein condensates are a flexible experimental platform to engineer synthetic quantum many-body systems. In particular, they host the so-called stripe phase, an instance of a supersolid state of matter. The peculiar excitation spectrum of the stripe phase, a definite footprint of its supersolidity, has been difficult to measure experimentally. Here, we perform in situ imaging of the stripes and directly observe both superfluid and crystal excitations. We investigate superfluid hydrodynamics and reveal a stripe compression mode, thus demonstrating that the system possesses a compressible crystalline structure. Through the frequency softening of this mode, we locate the supersolid transition point. Our results establish spin-orbit-coupled supersolids as ideal systems to investigate supersolidity and its rich dynamics.
\end{abstract}
\maketitle

Supersolidity is an exotic phase of matter characterized by two spontaneously-broken continuous symmetries, translational symmetry and U(1) gauge symmetry \cite{Boninsegni2012,Recati2023}. They are associated with two independent order parameters and manifest themselves, respectively, as a modulated density, reminiscent of the crystalline structure of a solid, and a global phase coherence, characteristic of a superfluid \cite{Andreev1969,Chester1970,Leggett1970}. In recent years, several quantum-gas platforms have realized superfluids with a spontaneous density modulation \cite{Leonard2017,Li2017,Tanzi2019, Boettcher2019, Chomaz2019}. Among these, Bose-Einstein condensates (BECs) with spin-orbit coupling, where the internal states of the particles are linked to their momentum through optical coupling \cite{Lin2011}, have been shown to host a supersolid state of matter where the density modulation takes the form of stripes. Owing to its conceptual simplicity, this so-called stripe phase is analytically tractable and serves as a powerful platform for exploring supersolid phenomena \cite{Wang2010,Ho2011,Lin2011,Li2012,Martone2021}. 

\begin{figure*}
\includegraphics[width=0.666\textwidth]{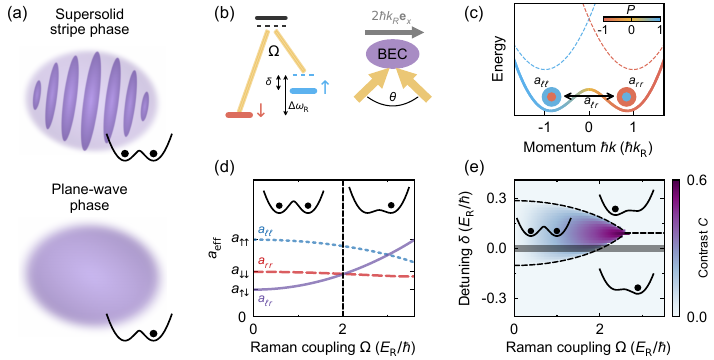}
\caption{\textbf{A robust high-contrast spin-orbit-coupled supersolid.} (a) Spin-orbit-coupled BECs at small coupling have a dispersion relation with two minima, leading to two possible many-body phases: the unmodulated plane-wave phase (one minimum occupied) and the modulated supersolid-stripe phase (both minima occupied). (b) Two-photon Raman coupling scheme with two beams of frequency difference $\Delta\omega_\mathrm{R}$, coupling strength $\Omega$, detuning $\delta$, and recoil momentum per beam $\hbar k_\mathrm{R}$ (tunable through the angle of the Raman beams $\theta$), used to engineer spin-orbit coupling. (c) Dispersion relation $E(k)$ of the lower optically-dressed state (solid line), compared to that of the bare atomic states (dashed lines). Colorscale: spin polarization $P$ of the dressed state. The system can be modeled as a mixture of condensates in the left $\ell$ and right $r$ wells, which consist of a coherent superposition of the bare spin states (red and blue circles) and have modified effective interaction properties given by the effective scattering lengths $a_{\ell\ell}$, $a_{\ell r}$ and $a_{rr}$. (d) Effective interaction parameters at $\delta=0$ vs. Raman coupling $\Omega$, calculated via the mixture model. Vertical black dashed line: supersolid phase transition point $\Omega_\mathrm{c}$ separating the supersolid-stripe phase and the plane-wave phase. This panel shows how tuning interaction strengths can stabilize the stripe phase vs. the plane-wave phase. (e) Phase diagram of a homogeneous spin-orbit-coupled BEC featuring a large and stable supersolid stripe phase, calculated with the variational ansatz of \cite{Li2012,Martone2015}. $C$ is the contrast of the density modulation. Dashed lines: phase boundaries calculated with the mixture model. Gray band: region experimentally explored in this work, with detuning $\hbar\delta=0.00\pm0.02E_\mathrm{R}$. Parameters in (d) and (e) correspond to $^{41}$K at a magnetic field $B=51.7$\,G and with a density $\SI{1e14}{\text{atoms}/\centi\meter^{3}}$, which realizes a robust and high-contrast supersolid stripe phase.}
\label{Fig:Fig1}
\end{figure*}

In supersolids, the existence of two order parameters leads to two gapless Goldstone modes and to a plethora of superfluid and crystal excitations \cite{Hofmann2021}. Probing this remarkably rich excitation spectrum has become a central topic of research across the different quantum-gas systems displaying supersolid properties \cite{LeonardScience2017,TanziNature2019,Guo2019,Natale2019,Tanzi2021,Norcia2022,Biagioni2024,Casotti2024}. Recently, the compressibility of the supersolids' emergent crystal structure has been put forward as an additional requirement for full-fledged supersolidity, given that solids naturally host crystal phonons. This property has been observed in the supersolid phase of dipolar quantum gases \cite{TanziNature2019,Natale2019,Norcia2022}, whereas it is lacking in BECs where effective long-range interactions are induced by single-mode cavities \cite{Leonard2017,LeonardScience2017}. Addressing this question in spin-orbit-coupled BECs has been challenging. Because of the extreme fragility of the realized stripe states, characterized by a vanishingly small stripe contrast \cite{Li2017,Putra2020}, the excitations of the supersolid phase in these systems have been considered experimentally inaccessible, and its properties have been widely debated \cite{Geier2023}. By utilizing spin-orbit-coupled potassium Bose-Einstein condensates, where the relative strengths of interspin and intraspin interactions are tunable through Feshbach resonances \cite{Tanzi2018}, we overcome the limitations of previous approaches and realize a stable stripe phase that persists at coupling regimes with strong modulation contrast. We are thus able to experimentally explore its excitation spectrum, show that its crystal structure is compressible, and precisely locate the supersolid phase transition. 

\section{Supersolid spin-orbit-coupled condensates} 
In spin-orbit-coupled BECs, supersolidity stems from the unconventional dispersion relation and the momentum-dependent interactions that arise thanks to the mixing of spin and spatial degrees of freedom \cite{Wang2010, Lin2011, Ho2011, Li2012}. The dispersion relation displays two minima in momentum space \cite{Higbie2004}, which results in two possible phases at the many-body level (Fig. \ref{Fig:Fig1}(a)). The plane-wave phase, which appears when a single minimum is occupied, has a homogeneous density profile. In contrast, the stripe phase, which occurs when both minima are occupied, has a spontaneously modulated density profile while it retains the spatial phase coherence of the condensate \cite{Wang2010, Ho2011, Lin2011, Li2012}, which constitutes a supersolid \cite{Hofmann2021}. The experimental challenge is to create a situation where the stripe phase occupies a significant portion of the phase diagram \cite{Lin2011}. To quantitatively investigate the stability of the stripe phase and find a suitable regime to probe supersolid excitations, we have developed an analytical model. 

This model considers the dressed states at the two minima of the dispersion relation, which emerges when two different states $\downarrow$ and $\uparrow$ (e.g., internal atomic states \cite{Lin2011, Putra2020}, or different bands of an optical lattice \cite{Li2017}) are coupled through an optical field that transfers momentum to the atoms. To realize this coupling, we use a pair of laser beams in a two-photon Raman configuration with frequency difference $\Delta\omega_\mathrm{R}$, two-photon Rabi frequency $\Omega$ and two-photon detuning $\delta$ (Fig. \ref{Fig:Fig1}(b)). It leads to a momentum transfer $2\hbar k_\mathrm{R}$ along the $x$ direction $\mathbf{e_x}$, with $\hbar k_{\mathrm{R}}$ the recoil momentum unit and $\hbar=h/(2\pi)$ the reduced Planck constant. The atom-photon dressed states have a momentum-dependent spin composition, which is set by the polarization parameter $P=\tilde{\delta}/\tilde{\Omega}$, where $\tilde{\delta}=\delta-2\hbar k_{\mathrm{R}} k/m$ and $\tilde{\Omega}=\sqrt{\Omega^2+\tilde{\delta}^2}$ are the generalized detuning and Rabi frequency. Here $m$ is the atomic mass and $\hbar k$ is the momentum along $x$ in the frame rotating at $\Delta\omega_\mathrm{R}$ and co-moving with the Raman coupling fields \cite{methods}. In this frame, the single-particle Hamiltonian of the system takes the translationally invariant form 
\begin{equation}
\mathcal{H}_{\mathrm{kin}}=\frac{\hbar^2}{2m} (\mathbf{k} + k_\mathrm{R} \sigma_z \mathbf{e_x} )^2 - \frac{\hbar \delta}{2} \sigma_z + \frac{\hbar \Omega}{2} \sigma_x, \label{H_SO} 
\end{equation}
where $\sigma_x$ and $\sigma_z$ are the Pauli matrices. For $\hbar\Omega<4\,E_\mathrm{R}$, where $E_\mathrm{R}=\hbar^2k_\mathrm{R}^2/2m$ is the Raman recoil energy, the dispersion relation of the lower dressed state features two minima (Fig. \ref{Fig:Fig1}(c)). The system behaves as a mixture of condensates in the left ($\ell$) and right ($r$) wells that are non-orthogonal. Thus, a modulated density profile emerges owing to an interference between them when both $\ell$ and $r$  are simultaneously occupied and miscible \cite{Higbie2004, Lin2011}. As the spin-orbit coupling strength increases, $\ell$ and $r$ become less orthogonal to each other. This larger wavefunction overlap leads to an increased contrast of the density modulation which, away from the transition, scales linearly with $\Omega$ \cite{methods}.

\begin{figure*}
\includegraphics[width=\textwidth]{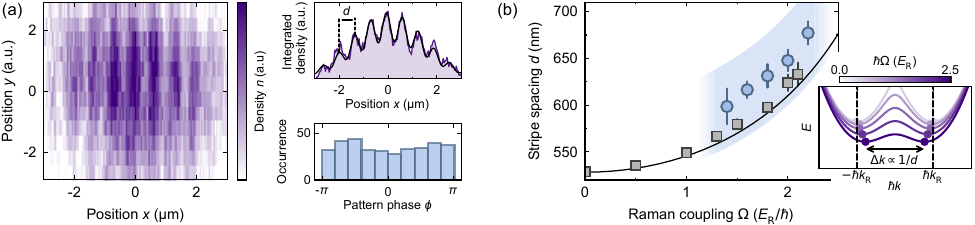}
\caption{\textbf{In situ observation of supersolid stripes.} (a) Single shot experimental image of the in situ density profile of the cloud at $\hbar\Omega=2.00\pm0.08\,E_\mathrm{R}$ measured with matter-wave optics. The image is rescaled to represent the expected cloud aspect ratio before matter-wave optics magnification along $x$. Top right: corresponding density profile integrated along the stripes. The black line denotes a modulated Gaussian fit with stripe spacing $d$. Bottom right: histogram showing the spatial phase of the modulation $\phi$ over 350 realizations. (b) Spacing $d$ of the stripes vs. Raman coupling $\Omega$ measured from in situ (blue circles) and time-of-flight (gray squares, see \cite{methods}). Solid line: single-particle theory prediction, obtained from the inverse of the momentum difference of the minima of the dispersion relation (shown in the inset for different spin-orbit coupling strengths $\Omega$)\ \cite{methods}. Errorbars correspond to one standard error of the mean (eom). The blue shaded area shows the systematic error of the in situ data from the magnification calibration.}
\label{Fig:Fig2}
\end{figure*}

Our effective field theory is based on the effective interactions of the optically-dressed states, which are given by the polarization at the dispersion minima \cite{methods}. We name it the mixture model because it builds on the mixture picture introduced in \cite{Higbie2004, Lin2011}. It delivers analytical expressions for both ground state properties and collective excitations \cite{methods} and highlights the conceptual simplicity of spin-orbit-coupled supersolids. Figure \ref{Fig:Fig1}(d) shows the effective interaction parameters according to our mixture model for a gas with bare scattering lengths $a_{\uparrow\uparrow}$, $a_{\downarrow\downarrow}$, and $a_{\uparrow\downarrow}$ at detuning $\delta=0$. Increasing $\Omega$ reduces the effective intra-well interactions $a_{\ell\ell}$, $a_{rr}$ and increases the inter-well ones $a_{\ell r}$. As a result, the miscibility of the system is reduced and it becomes energetically favorable to occupy a single minimum of the dispersion relation as soon as $a_{\ell r}>a_{\ell\ell}$ or $a_{\ell r}>a_{rr}$. The crystal-like modulation is then lost and the system transitions to the plane-wave state. This relation between interactions and coupling strength demonstrates that obtaining a stripe phase at large couplings requires $a_{\uparrow\downarrow}\ll a_{\uparrow\uparrow},a_{\downarrow\downarrow}$. This situation is not available in atomic species like the $^{87}$Rb used in previous works, where the small scattering length difference yields a very fragile supersolid phase with a contrast of the density modulation  $C<5\%$ \cite{Lin2011,Ji2014,Bersano2019, Putra2020}. However, we can achieve it by exploiting two-component potassium BECs, where the interspin interactions are tunable using a Feshbach resonance \cite{Tanzi2018}. The complete phase diagram of the system with $^{41}$K at 51.7\,G in the parameter space of $\Omega$ and $\delta$ is depicted in Fig. \ref{Fig:Fig1}(e). The stripe phase appears as a dome that is not centered at $\delta=0$ owing to the setting $a_{\downarrow\downarrow}<a_{\uparrow\uparrow}$. The phase diagram shows the contrast $C$ according to the variational model of Refs. \cite{Li2012,Martone2015}, in excellent agreement with the black dashed lines, which are analytical predictions from our mixture model. This constitutes an unusually large, stable, and high-contrast supersolid stripe phase and provides an excellent regime in which to perform our experiments.

\section{In situ observation of supersolid stripes} We realize the supersolid stripe phase with a harmonically trapped  Bose-Einstein condensate of $^{41}$K atoms in states $\ket{\downarrow} \equiv \ket{F=1,m_F=0}$ and $\ket{\uparrow} \equiv \ket{F=1,m_F=-1}$. By setting the magnetic field to $B=51.7\pm 0.1$\,G, between an intra-spin and an inter-spin Feshbach resonance, we tune the bare state interactions to $(a_{\uparrow\uparrow},a_{\downarrow\downarrow},a_{\uparrow\downarrow})=(115\pm10,65\pm1,40\pm8)\,a_0$, where $a_0$ is the Bohr radius \cite{Tanzi2018}. We couple the two states with two Raman laser beams of orthogonal polarizations at the tune-out wavelength $\lambda_\mathrm{R} = 768.97$\,nm, for which scalar potentials cancel. We work at a zero two-photon Raman detuning $\hbar\delta=0.00\pm0.02\,E_\mathrm{R}$ (gray band in Fig. \ref{Fig:Fig1}(e)), where the phase transition point is independent of the atomic density for our parameter regime \cite{Chisholm2023}. The Raman beams have a relative angle of $\theta = 94\pm1^\circ$, which leads to a momentum transfer along the $x$ direction of $2\hbar k_\mathrm{R} = 4\pi\hbar\mathrm{sin}(\theta/2)/\lambda_\mathrm{R}$ (Fig. \ref{Fig:M1}) \cite{methods}. See Tab. \ref{tab:exp_params} \cite{methods} for details of the experimental parameters used throughout the text. 

A major challenge for the observation of the supersolid density modulation is the large optical density of the cloud and the small period of the stripes $d$ ($<$ \SI{1}{\micro\meter}). Previous experiments circumvented this difficulty by indirectly revealing the density modulation through Bragg scattering of light \cite{Li2017,Putra2020}. We instead obtain direct access to the stripes by magnifying the density distribution by a factor of $25\pm2$ along the modulated direction with matter-wave optics, before observing the magnified stripes with absorption imaging (Fig. \ref{Fig:M3}) \cite{methods}. Figure \ref{Fig:Fig2}(a) shows a typical in situ atomic density distribution, as well as the corresponding integrated density profile for a spin-orbit coupling strength $\hbar\Omega=2.00\pm0.08\,E_\mathrm{R}$. The presence of harmonic confinement favors a supersolid modulated density profile with a maximum at the center of the trap. However, for our experimental parameters the energy penalty for displacing the stripes is negligible (Fig. \ref{Fig:M2}) \cite{methods}. As a result, the stripe position, given by the spatial phase $\phi$ of the modulation pattern, takes random values from one experimental realization to another (Fig. \ref{Fig:Fig2}(a), bottom right).

To investigate the periodicity of the stripes, we extract the stripe spacing from in situ images taken at different spin-orbit coupling strengths $\Omega$. Figure \ref{Fig:Fig2}(b) shows that the stripe spacing $d$ increases significantly with $\Omega$, which can be understood from a momentum-space picture. The stripe spacing is given by the momentum difference $\Delta k$ of the two dressed states through $d=2\pi/\Delta k$  which, at equilibrium, corresponds to the difference between the two minima of the dispersion relation (Fig. \ref{Fig:Fig2}(b), inset). It can also be inferred from spin-resolved momentum-space images obtained through time-of-flight expansion (Fig. \ref{Fig:M4}) \cite{methods}. Contrary to the in situ analysis, the analysis of the time-of-flight data is model-dependent, resulting in a spacing $\pi/k_\mathrm{R}$ at the zero-coupling point by construction. The data are shown as gray squares in Fig. \ref{Fig:Fig2}(b). Within their respective uncertainties, both in situ and time-of-flight data agree with the theory prediction. Thus, our observations demonstrate that the period of spin-orbit-coupled supersolids is not externally imposed by the wavelength of the coupling laser. This variable crystalline pattern stands in contrast to that of cavity systems, where the modulation period is fixed by the light wavelength \cite{LeonardScience2017}.

\section{Probing superfluid hydrodynamics through excitations}
\begin{figure}
\includegraphics[width=0.333\textwidth]{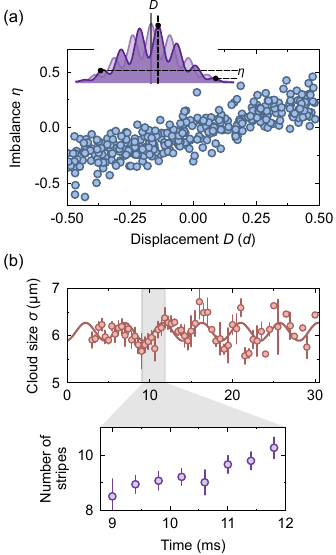}
\caption{\textbf{Crystal Goldstone mode and superfluid hydrodynamics.} 
(a) Population imbalance between side stripes $\eta$ vs. displacement of the central stripe $D$ with respect to the center of mass of the cloud (see inset and \cite{methods}). For a harmonically-trapped system with superfluid flow, where a sliding of the density modulation is compensated by particle exchange between the stripes, the two quantities are linearly correlated and provide evidence for the crystal Goldstone mode. (b) Top panel: breathing mode of the overall cloud, observed through its size $\sigma$ vs. time. Solid line: sinusoidal fit used to extract the breathing frequency $\omega_{\mathrm{B}}$, yielding $\omega_{\mathrm{B}}/\omega_{\mathrm{D}}=1.5\pm0.1\sim\sqrt{5/2}$ with $\omega_{\mathrm{D}}$ the dipole mode frequency, as expected from superfluid hydrodynamics. Bottom panel: number of stripes in the supersolid cloud. It dynamically changes during half of a breathing oscillation (see shaded area in top panel), supporting the existence of superfluid flow across the system. Errorbars denote one standard eom.}
\label{Fig:Fig3}
\end{figure}

Supersolids are characterized by a superfluid flow through the crystalline structure that maintains the global phase coherence of the system \cite{Leggett1970,Boninsegni2012}. Whereas previous experiments demonstrated phase coherence in spin-orbit-coupled supersolids using Talbot interferometry \cite{Putra2020}, our in situ images allow us to investigate instead the superfluid hydrodynamic properties of the system. To this end, we followed the procedure developed for supersolids in dipolar gases \cite{Guo2019} and studied the population imbalance between stripes. Figure \ref{Fig:Fig3}(a) shows the imbalance between side stripes $\eta$ as a function of the displacement $D$ of the central stripe with respect to the center of mass of the cloud (see illustration of $D$ and $\eta$ in the inset and \cite{methods}). We see a clear linear correlation between displacement and imbalance, revealing that a phase shift of the crystal structure is compensated by a superfluid counterflow. Such a phase shift provides evidence that the low-energy crystal Goldstone mode of the finite-size system is excited \cite{Guo2019}.

To probe other signatures of superfluidity, we investigate the collective breathing and dipole modes of the system, which we detect through the evolution of the cloud's width $\sigma$ and the displacement of its center of mass $x_\mathrm{0}$, respectively. In conventional BECs, a ratio of their frequencies $\omega_{\mathrm{B}}/\omega_{\mathrm{D}}=\sqrt{5/2}$ (with $\omega_\mathrm{B}$ and $\omega_\mathrm{D}$ the breathing and dipole mode frequencies) signals a system described by superfluid hydrodynamics and thus supports the superfluid character of the cloud \cite{Menotti2002}. Figure~\ref{Fig:Fig3}(b) shows the breathing oscillation of our system. We find a frequency of $\omega_\mathrm{B}/\omega_\mathrm{D}= 1.5\pm0.1$, in good agreement with the conventional BEC prediction. This behavior is consistent with numerical simulations of the Gross-Pitaevskii equation which implicitly assume superfluid hydrodynamics \cite{Chisholm2023}. Here the dipole mode frequency $\omega_{\mathrm{D}}$ differs from the trapping frequency $\omega_x$ because spin-orbit-coupled systems violate Kohn's theorem \cite{Li2012b,Zhang2012}, analogously to superfluids subjected to an optical lattice \cite{Kraemer2002,Fort2003}. Another notable aspect is that, owing to the changing size of the cloud, the number of stripes in the supersolid varies with time. The bottom panel of Fig. \ref{Fig:Fig3}(b) shows that the superfluid flow can dynamically create additional stripes. Thus, our data support the superfluid character of the system via the frequency of its breathing mode and the existence of dynamic particle flow between the stripes.

\begin{figure*}
\includegraphics[width=\textwidth]{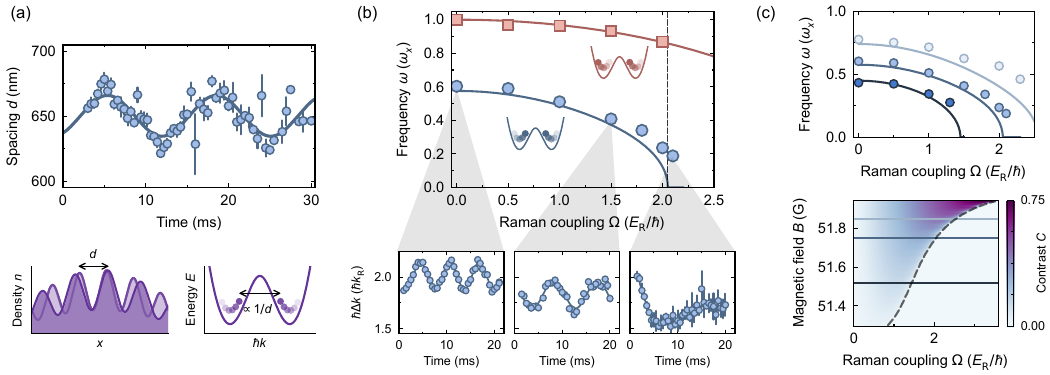}
\caption{\textbf{Stripe compression mode and supersolid phase transition. }
(a) In situ observation of the stripe compression mode. After a rapid ramp of the coupling strength $\Omega$ ending at $t=0$, the spacing of the stripes $d$ oscillates in time, demonstrating that the crystal structure is not stiff. Errorbars denote one eom. Bottom: sketch of the stripe compression in real space (left) and in momentum space (right). (b) Observation of mode softening from momentum-space measurements. Top: collective frequencies of the dipole mode (red squares) and stripe compression mode (blue circles), in units of the trapping frequency $\omega_x$, vs. coupling strength $\Omega$. Solid lines: mixture model predictions without fitting parameters. Vertical black dashed line: thermodynamic limit prediction of the critical point $\Omega_\mathrm{c}$ \cite{methods}. Bottom: example oscillations of the relative momentum of the dressed states $\hbar\Delta k$ for $\hbar\Omega=0$, $1.50\pm0.06$ and $2.10\pm0.08\,E_\mathrm{R}$, with the exponentially decaying sinusoidal fits (solid lines) used to extract the corresponding frequencies. Errorbars: error on the fit (top panel) and standard deviation of up to three repetitions (bottom panel). (c) Top: frequency softening of the stripe compression mode for magnetic fields $B= 51.52\pm0.02$, $51.75\pm0.02$ and $51.85\pm0.02$\,G (from dark to light blue, respectively). Bottom: phase diagram of the supersolid-to-plane-wave phase transition, calculated from the mixture model. Solid blue lines: magnetic field values corresponding to the top panel. Gray dashed line: phase boundary predicted by the mixture model.}
\label{Fig:Fig4}
\end{figure*}

\section{Stripe compression excitations and location of the supersolid phase transition} 
Supersolids host a rich excitation spectrum originating from both their crystal-like and their superfluid character. Previous studies have revealed significant differences between platforms. Dipolar supersolids host crystal phonon-like modes \cite{TanziNature2019, Natale2019}. Instead, phonons are absent in cavity systems and multimode cavities are required to restore them \cite{Guo2021}. This difference has led to controversial discussions on the stiffness of the stripe pattern of spin-orbit-coupled systems, which were only recently clarified theoretically with the prediction of stripe compression modes akin to those of dipolar supersolids \cite{Geier2023}. Here, we address this question experimentally. To this end, we prepare the system by a non-adiabatic ramp of the spin-orbit coupling strength $\Omega$, which is slow enough to remain in the lower dressed state, but sufficiently fast to excite the stripe pattern \cite{methods}. Figure \ref{Fig:Fig4}(a) depicts the evolution of the modulation period $d$ over time for a system with $\hbar\Omega=2.00\pm0.08\,E_\mathrm{R}$. Notably, we observe that $d$ is not constant, but instead displays a dynamical oscillation corresponding to a stripe compression mode. This proves that the stripe phase of a spin-orbit-coupled BEC has a compressible crystal structure. Observing this mode is a signature of full-fledged supersolidity, as it reveals the existence of a crystal phonon branch \cite{Geier2023}.  

In a momentum-space picture, this mode is caused by an out-of-phase oscillation of the $\ell$ and $r$ condensates in their dispersion minima. Within our effective mixture model, it corresponds to a spin-dipole mode of the dressed states. Owing to its spin nature, it decouples from collective modes that involve instead the density degree of freedom, such as the breathing mode observed in Fig. \ref{Fig:Fig3}(b) \cite{methods}. This mode disappears at the transition to the plane-wave phase, although it can be observed in out-of-equilibrium configurations \cite{Li2019}. Its frequency has been theoretically predicted to reduce when approaching the supersolid stripe to plane-wave phase transition, reaching a minimum at the transition point $\Omega_\mathrm{c}$ \cite{Chen2017, Geier2021}. In the next series of experiments, we therefore focus on determining the critical value $\Omega_\mathrm{c}$ for a spin-orbit coupling detuning $\delta=0$ through the frequency of the compression mode. To measure its softening, it is essential to prepare the ground state of the system. In spin-orbit-coupled BECs, this is challenging because of the existence of very long-lived metastable states  \cite{Lin2011} with spin polarization $P\not=P_0$, where $P_0$ is the polarization of the ground state. Such metastable states make it difficult to detect the phase transition point based only on the contrast of the stripes. Instead, in the limit of large system sizes our mixture model predicts that, for each value of $\Omega$, $P_0$ minimizes the stripe compression mode frequency $\omega_\mathrm{SC}$ \cite{methods}. We thus experimentally minimize $\omega_\mathrm{SC}$ as a function of polarization $P$ for different coupling strengths $\Omega$, which allows us to investigate the expected mode softening of the compression mode when approaching the critical point $\Omega_\mathrm{c}$ (Fig. \ref{Fig:M5}) \cite{methods}. To enhance our detection in the regimes of weak coupling $\Omega$ or large spin polarization $P$, we exploit momentum space images to perform this measurement. 

Figure \ref{Fig:Fig4}(b) summarizes our results. It displays the frequencies of the dipole and stripe compression modes as a function of the spin-orbit coupling strength $\Omega$, measured at the polarization $P$ that experimentally minimizes $\omega_\mathrm{SC}$, thus approximating the ground state polarization $P_0$. We observe that the frequency of both modes diminishes when increasing $\Omega$. For the dipole mode (red squares), this is caused by the reduction of curvature of the dispersion relation at the minima (i.e., an increase of the effective mass \cite{Lin2011b, Khamehchi2017}). The frequency of the dipole mode vanishes at the plane-wave to single-minimum transition, which in the single-particle regime takes place at $\hbar\Omega=4\,E_R$ and has been investigated in previous works \cite{Li2012b, Zhang2012}. It does not show particular features in the Raman coupling range explored here. In contrast, the frequency of the compression mode (blue circles), dramatically decreases when approaching the expected supersolid phase transition point $\Omega_\mathrm{c}$ (vertical black dashed line).
The softening of the mode can also be clearly observed in the three representative oscillations included in the figure. 

We compare our experimental results with our theoretical prediction (solid lines) and find a good agreement. The theoretical lines have no fitting parameters and result from a method originally developed to analytically predict the mode frequencies in conventional mixtures \cite{Cavicchioli2022}, which we adapt to our mixture model \cite{methods}. In the vicinity of the phase transition point the experimental data deviates from the thermodynamic-limit theory because of finite size effects, which give a finite gap to the zero-energy modes. All measurements displayed in Fig. \ref{Fig:Fig4}(b) are taken at a magnetic field $B=51.75\pm0.02$\,G, resulting in a supersolid phase transition located at $\hbar\Omega_\mathrm{c}=2.05\pm0.08\,E_{\mathrm{R}}$ (vertical dashed line). By exploiting the rich Feshbach spectrum of $^{41}$K \cite{Tanzi2018}, we can furthermore tune the bare-state interactions and displace the phase transition point. The top panel of Fig. \ref{Fig:Fig4}(c) summarizes the results of such measurements, which are in good agreement with the theoretical expectations (solid lines) and the predicted phase diagram of the system as a function of the magnetic field (bottom panel). This final measurement showcases the tunability of the supersolid phase transition in our system. 

\section{Discussion and outlook}  
Our realization of a high-contrast supersolid stripe phase in a spin–orbit-coupled BEC and the in situ observation of its crystal modes enables the observation of other excitations, in particular those involving the spin degree of freedom such as collective spin breathing \cite{Chen2017,Chisholm2023}, stripe angle oscillations \cite{Geier2023}, and the Higgs amplitude mode of the translational symmetry breaking \cite{LeonardScience2017}. Although the existence of a crystal Higgs mode is a fundamental property of supersolids, in dipolar systems it remains unobserved because it couples to higher excited modes \cite{Hertkorn2019,Hertkorn2024}. In contrast, our mixture model suggests that in spin-orbit-coupled supersolids, the spin character of the Higgs mode keeps it decoupled from density excitations, making it more accessible.  Besides their fundamental interest, investigating such excitations would clarify how the spin degree of freedom distinguishes spin-orbit-coupled supersolids from other platforms. Our system also opens the possibility of investigating the superfluid fraction across the supersolid phase transition \cite{Zhang2016}, in analogy to very recent experiments with dipolar gases \cite{Biagioni2024}, but closer to the thermodynamic limit. Finally, by setting the effective interactions between the dressed condensates to attractive values ($a_{\ell r}<0$), it should be possible to cancel the overall mean-field energy and reveal beyond mean-field effects. These are expected to stabilize quantum liquid droplets \cite{Cabrera2018, Semeghini2018} with supersolid properties \cite{Sachdeva2020,Sanchez-Baena2020}, with lifetimes long enough to be observed. The ability to independently control the magnitude of beyond mean-field effects is a unique feature of spin-orbit-coupled supersolids, which should allow disentangling the effects of quantum fluctuations and supersolidity.

\section*{Acknowledgments}
We acknowledge insightful discussions with S. Stringari, K. T. Geier, G. I. Martone, P. Hauke, W. Ketterle, I. B. Spielman, F. Ferlaino, and T. Donner. We also wish to thank T. Donner, W. Ketterle, P. Massignan, A. Rubio-Abadal, and S. Stringari for a critical reading of the manuscript.

\paragraph*{Funding:}
We acknowledge funding from the European Union (ERC CoG-101003295 SuperComp), the Spanish Ministry of Science and Innovation MCIU/AEI/10.13039/501100011033 (projects LIGAS PID2020-112687GB-C21, MAPS PID2023-149988NB-C22, and Severo Ochoa CEX2024-001490-S at ICFO and projects LIGAS PID2020-112687GB-C22 and MAPS PID2023-149988NB-C21 PID2020112687GB-C22 at UAB), the EU QuantERA project DYNAMITE (funded by MICN/AEI/ 10.13039/501100011033 and by the European Union NextGenerationEU/PRTR PCI2022-132919 (Grant No. 101017733)), Deutsche Forschungsgemeinschaft (Research Unit FOR2414, Project No. 277974659), Generalitat de Catalunya (AGAUR SGR 2021-SGR-01448 at ICFO and 2021-SGR-00138 at UAB, CERCA program), Fundació Cellex and Fundació Mir-Puig.
C.S.C., S.H., and R.R. acknowledge support from the European Union (Marie Skłodowska-Curie–713729, Marie Skłodowska Curie–101149245 Epiquant, and Marie Skłodowska-Curie–101030630 UltraComp). V.B.M. acknowledges support from the Beatriu de Pinós Program and the Ministry of Research and Universities of the Government of Catalonia (Grant No. 2019-BP-00228). J.C. acknowledges support from the Ministry of Economic Affairs and Digital Transformation of the Spanish Government through the QUANTUM ENIA project call – Quantum Spain project, and A.C. acknowledges support from the UAB Talent Research program during the early stages of this work. 

\paragraph*{Author contributions:}
C.S.C. and R.R. developed and took preliminary data on the first Raman coupling and matter-wave optics magnification experimental schemes. S.H., V.B.M. and R.V. developed the final magnification scheme, devised the experimental collective excitation methods and observables, and took the final data. S.H. and R.V. analyzed the final data. J.C., C.S.C, L.T., and A.C. developed the theory. J.C., C.S.C., and V.B.M. performed numerical simulations. S.H., R.V., J.C., and L.T. wrote the manuscript. L.T. developed the concept, and A.C. and L.T. supervised the work.

\paragraph*{Competing interests:}
There are no competing interests to declare.

\paragraph*{Data and materials availability:}
The data shown in the main text and extended data figures are available from the CORA RDR repository \cite{Data}.

\renewcommand{\thefigure}{S\arabic{figure}}
\renewcommand{\thetable}{S\arabic{table}}
\renewcommand{\theequation}{S\arabic{equation}}
\renewcommand{\thepage}{S\arabic{page}}

\setcounter{figure}{0}
\setcounter{table}{0}
\setcounter{equation}{0}
\setcounter{section}{0}

\clearpage

\begin{center}
\section*{\large\textbf{Materials and Methods}} 
\end{center}

\section{Experimental parameters and system preparation}

\begin{figure}[b]
\includegraphics{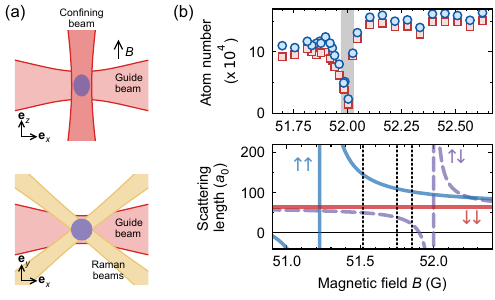}
\caption{\textbf{Experimental setup and scattering length.} (a) Sketch of the optical setup. Top panel: $x$-$z$-plane (side view) with confining beam, guide beam and direction of the magnetic field $B$. Bottom panel: $x$-$y$-plane (top view) with guide beam and Raman beams. (b) Top panel: location of the $\uparrow\downarrow$ Feshbach resonance through loss spectroscopy. Blue circles (red squares) show the atom number of state $\uparrow$ ($\downarrow$). The shaded area indicates the field $B_0$ of the resonance. Bottom panel: dependence of the $a_{\uparrow\downarrow}$ (purple dashed line), $a_{\uparrow\uparrow}$ (blue solid line) and $a_{\downarrow\downarrow}$ (red solid line) scattering lengths on the magnetic field $B$ as resulting from the loss spectroscopy measurements and \cite{Tanzi2018}. Vertical black dotted lines line: magnetic field values 
used in this study.
}
\label{Fig:M1}
\end{figure}

\textit{BEC preparation.} We perform the experiments with a $^{41}$K BEC in the Zeeman sublevels of the $F=1$ hyperfine manifold $\ket{\downarrow} \equiv \ket{F=1,m_F=0}$ and $\ket{\uparrow} \equiv \ket{F=1,m_F=-1}$. To prepare an incoherent spin mixture of variable spin composition, we start with a BEC in $\ket{\downarrow}$, apply a radio-frequency (rf) pulse to transfer the cloud partially to state $\ket{\uparrow}$, and hold the cloud for \SI{3}{ms} to let it decohere. The spin composition of the system depends on the length of the rf pulse. Throughout the paper, we quantify it via the polarization parameter of the cloud $P$, which we experimentally measure as $P=(N_\uparrow-N_\downarrow)/(N_\uparrow + N_\downarrow)$ with $N_i$ the number of atoms in state $i=\uparrow,\downarrow$. 

\begin{figure}
\includegraphics{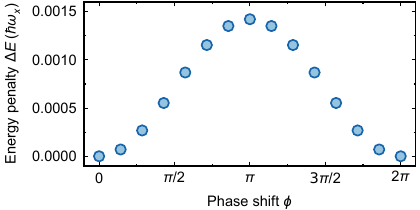}
\caption{\textbf{Quantifying the explicit symmetry breaking induced by the confining potential. }
The energy penalty $\Delta E$ of a spatial phase shift $\phi$ of the modulation pattern is negligible compared to the energy of the harmonic trap along the modulated direction $\hbar\omega_x$. The experimental parameters correspond to those of Figure \ref{Fig:Fig2}(a) of the main text, i.e., a Raman coupling strength $\hbar\Omega=2E_\mathrm{R}$ and the in situ scattering lengths, atom number and trapping frequencies of table \ref{tab:exp_params}.}
\label{Fig:M2}
\end{figure}

\textit{Confining potentials.} We trap the atoms in a crossed optical dipole trap consisting of a guide beam along the $x$ direction and a confining beam along the $z$ direction, see Figure \ref{Fig:M1}(a). It creates a spin-independent potential on the atoms $V_{\mathrm{ho}}=m(\omega_x^2x^2+\omega_y^2y^2+\omega_z^2z^2)/2$, where $\omega_i$ are the trapping frequencies, see table \ref{tab:exp_params}. As discussed in the main text, the harmonic trapping potential creates a small energy penalty for the displacement of the supersolid stripes. Figure \ref{Fig:M2} shows its computed value for the parameters of Figure \ref{Fig:Fig2}(a) (in situ values of table \ref{tab:exp_params}). The maximum value of the energy penalty is $\sim0.2$\,Hz, which can be easily overcome by other processes such as thermal excitations or non-adiabaticities in the preparation of the system.

\begin{table}
    \centering
    \begin{tabular}{ p{2.2cm}  p{2cm} p{4cm} }
    \hline
       Parameter  & In situ & Time-of-flight \\
       \hline
        $\omega_x/2\pi$ (Hz) & $150\pm7$ & $270\pm12$\\
        $\omega_y/2\pi$ (Hz) &  $180\pm15$ & $290\pm15$\\
        $\omega_z/2\pi$ (Hz) & $103\pm5$ & $103\pm5$\\
        \hline
        $B$ (G) & $51.7\pm0.1$ & $51.52\pm0.02$, $51.75\pm0.02$,\\ &&$51.85\pm0.02$\\
        \hline
        $a_{\uparrow\uparrow}$ ($a_0$) & $115\pm10$ & $146\pm6$, $110\pm2$, $103\pm1$\\
        $a_{\downarrow\downarrow}$ ($a_0$) & $65\pm1$ & $65\pm1$, $65\pm1$, $65\pm1$\\
        $a_{\uparrow\downarrow}$ ($a_0$) & $40\pm8$ & $50\pm1$, $39\pm2$, $23\pm6$\\
        \hline
        $N$ ($10^3$) & typ. 15 & $35\pm5$\\
    \hline
    \end{tabular}
    
    \caption{\textbf{Experimental parameters.} Trapping frequencies in the crossed optical dipole trap ($\omega_x$,$\omega_y$, $\omega_z$), magnetic field $B$, scattering lengths $a_{\uparrow\uparrow}$, $a_{\downarrow\downarrow}$, and $a_{\uparrow\downarrow}$ in units of the Bohr radius $a_0$, and total atom number $N$, for the different experiments presented in the main text. In situ experiments: Figures \ref{Fig:Fig2}, \ref{Fig:Fig3}, and \ref{Fig:Fig4}(a). Time-of-flight experiments: Figures \ref{Fig:Fig2}(b), \ref{Fig:Fig4}(b) and \ref{Fig:Fig4}(c).}
    \label{tab:exp_params}
\end{table}

\textit{Interactions.} We set the magnetic field $B$ in the range $51.5-51.9$\,G, in between two Feshbach resonances that allow us to adjust the bare scattering lengths $a_{\uparrow\uparrow}$ and $a_{\uparrow\downarrow}$, leaving $a_{\downarrow\downarrow}$ unchanged \cite{Tanzi2018}. To improve the accuracy of the $a_{\uparrow\downarrow}$ scattering length values, to which the stripe compression mode frequency is very sensitive, we measure the location of its Feshbach resonance through loss spectroscopy in our BEC mixture (see Fig. \ref{Fig:M1}(b)). To this end, we ramp the magnetic field from 54.2\,G to a variable value $B$ within \SI{0.5}{\milli\second} and hold the cloud for \SI{15}{\milli\second}. The losses are maximal at $B_0=52.00\pm0.03$\,G, which we identify with the position of the resonance. The scattering length vs. magnetic field values used throughout the paper are depicted in Figure \ref{Fig:M1}(b) (bottom panel), where we have shifted the predictions of $a_{\uparrow\downarrow}$ from \cite{Tanzi2018} according to our new calibration. The interaction parameters used for the different measurements are summarized in table \ref{tab:exp_params}. There, errorbars in the scattering lengths correspond to the field stability during the measurements and do not take into account potential systematic errors on the Feshbach resonance parametrization \cite{Tanzi2018}.

\textit{Raman coupling.} We introduce spin-orbit coupling along the $x$ axis by coupling the states $\ket{\downarrow}$ and $\ket{\uparrow}$ with a two-photon Raman transition of two-photon Rabi frequency $\Omega$ and detuning $\delta$, which we calibrate as in our previous work \cite{Froelian2022}. For all measurements, we set $\delta=0$, where the phase boundary is independent of the atomic density for the densities we realize in the experiment. The Raman lasers have orthogonal polarizations and operate at the tune-out wavelength $\lambda_\mathrm{R} =$ 768.97\,nm \cite{Jiang2013a,Jiang2013b}, thus avoiding additional confinement. After preparing the spin mixture, the spin-orbit coupling strength $\Omega$ is ramped within $t_\mathrm{ramp}=\SI{30}{ms}$ (\SI{6}{ms} for time-of-flight data) from zero to its final value. This ramp is slow enough to avoid excitations to the higher Raman dressed band ($2\pi/\Omega\approx\SI{0.1}{ms}$), but fast enough to excite collective modes of the stripes (in situ: $2\pi/\omega_x\approx\SI{7}{ms}$, time-of-flight: $2\pi/\omega_x\approx\SI{4}{ms}$). To excite the stripe compression mode at low Raman coupling $\hbar \Omega<1\,E_\mathrm{R}$ (data points in Fig. \ref{Fig:Fig4}), we first ramp the coupling to $1\,E_\mathrm{R}$ and then reduce it to its final value.

\section{In situ experiments}

\begin{figure}
\includegraphics{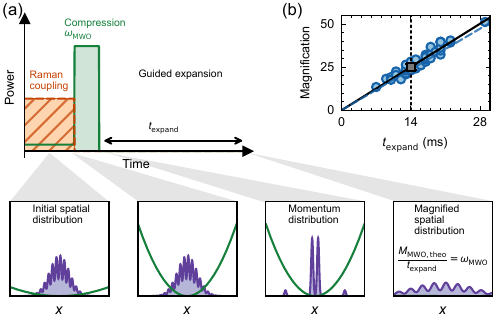}
\caption{\textbf{Matter-wave optics magnification.}
(a) Experimental magnifying sequence, which consists of two phases: the compression phase and the guided expansion phase. Green area: power of the optical confining beam. Orange dashed area: power of the Raman coupling beams. Insets: sketches of the density distribution of the cloud at different key steps of the sequence (purple). The solid green line represents the harmonic trap made by the confining beam. 
(b) Calibration of the magnification of the matter-wave optics scheme. The circles correspond to the measured magnification as a function of expansion time $t_{\mathrm{expand}}$. Solid line: linear fit. Black square: resulting magnification for the guided expansion time used in the main text ($t_\mathrm{expand}=14$\,ms, dotted line). Dashed line: theoretical expectation $M_\mathrm{MWO,theo}=\omega_\mathrm{MWO}t_\mathrm{expand}$.}
\label{Fig:M3}
\end{figure}

\textit{Matter-wave optics scheme.} To resolve in situ the supersolid stripes, we adapt previous matter-wave optics methods to magnify the density profile of the cloud \cite{Shvarchuck2002,Murthy2014,Asteria2021} to a spin-orbit-coupled system. Our scheme magnifies the profile only along the direction of the stripes $x$ and consists of several steps summarized in Figure \ref{Fig:M3}. First, we simultaneously turn off the Raman coupling (projecting the dressed states back into the bare states) and increase the confinement of the harmonic trap along the $x$ direction to a frequency $\omega_\mathrm{MWO}$. Then, we let the atoms evolve in the trap for a quarter of the $x$ trap period $T=2\pi/\omega_\mathrm{MWO}$. Subsequently, we switch the $x$ confinement off and let the cloud expand in a guide beam along $x$ for a time $t_\mathrm{expand}=14$\,ms, during which we switch off the magnetic field. This leads to a magnification of the in situ density profile by a factor $M_\mathrm{MWO,theo}=\omega_\mathrm{MWO}t_\mathrm{expand}=24\pm1$, see experimental calibration below. Finally, we perform spin-independent absorption imaging at zero magnetic field to obtain the total density profile of the cloud $n=n_\uparrow+n_\downarrow$ and observe the magnified stripes. Due to the guiding beam, our scheme magnifies the density distribution only along $x$ and not along $y$. We expect a small effect on the cloud size in $y$ direction from the sudden increase of power of the confining beam, which is why the $y$ pixelsize in Figure \ref{Fig:Fig2}(a) is indicated in arbitrary units (a.u.). 

\textit{Matter-wave optics calibrations.} We calibrate the magnification of the matter-wave optics sequence by applying a standing-wave optical lattice along the $x$ direction on the BEC spin-mixture in the absence of Raman coupling. Its spacing is $d_\mathrm{calib}=\lambda_\mathrm{calib}/2$, where $\lambda_\mathrm{calib}=1064$\,nm. We apply our magnifying scheme described above for guided expansion times $t_\mathrm{expand}$ ranging between $7$ to $29$\,ms and extract the magnification from a linear fit to the data, see Figure \ref{Fig:M3}(c). We obtain a magnification of $M_\mathrm{MWO}=25\pm2$, which is compatible with the theory expectation $M_\mathrm{MWO,theo}=24\pm1$.

To verify the effect of interactions during the matter-wave optics sequence, we create a moving lattice and exploit Bragg diffraction of a single component BEC. Pulsing the lattice imparts momentum to a fraction of the atoms that depends on the Bragg pulse length $t_{\mathrm{Bragg}}$. The interference between the diffracted and undiffracted atoms then yields a modulated density profile of spacing $d_{\mathrm{calib}}$, whose contrast can be controlled through $t_{\mathrm{Bragg}}$. By comparing the expected and measured contrast as a function of the pulse time, we conclude that interactions lead to scattering halos that significantly reduce the contrast of the magnified modulated density profiles, but do not affect the spacing of the modulation \cite{Chisholm2023}. In the supersolid data, we thus restrict our analysis to the stripe spacing and do not exploit the measured values of the contrast.

\textit{Data analysis.} 
We post select our in situ data by discarding images if the Fourier transform of the integrated 1D density profile does not feature a peak above a fixed threshold in the spatial frequency region of interest ($>\SI{0.05}{\micro\meter^{-1}}$). 37\% of the images where kept for Figure \ref{Fig:Fig2}(b) and 94\% of the images where kept for Figures \ref{Fig:Fig2}(b), \ref{Fig:Fig3} and \ref{Fig:Fig4}(a). This difference comes from varying day-to-day experimental conditions. Since the stripe contrast scales linearly with the spin-orbit coupling strength $\Omega$, stripe spacing measurements were limited to $\Omega>1.4\,E_\mathrm{R}/\hbar$. Following the theoretical analysis of \cite{Li2012}, the integrated 1D density profile is then fitted by a modulated Gaussian 
\begin{equation}
           N_\mathrm{fit}(x)=A\mathrm e^{-(x-x_\mathrm{0})^2/2\sigma^2}\times\left[1+C\sin\left(\frac{2\pi (x-x_\mathrm{0})}{d}+\phi\right)\right].
     \label{eq:modGauss}
\end{equation}
Here, $d$, $\phi$ and $C$ are respectively the spacing, the phase and the contrast of the supersolid stripe pattern, while $x_\mathrm{0}$ and $\sigma$ are the center-of-mass position and the characteristic size of the cloud. The initial guess for the spacing is obtained from the position of the peak in the Fourier analysis. Such a fit is shown in Figure \ref{Fig:Fig2}(a). All the in situ data presented in Figures \ref{Fig:Fig2}, \ref{Fig:Fig3} and \ref{Fig:Fig4}(a) is extracted from the fitting parameters yielded by this procedure. In Figure \ref{Fig:Fig3}(a), we calculate the imbalance $\eta$ as the normalized difference of the integrated density of the third stripe to the left and the third stripe to the right of the center stripe, whose positions we obtain from the fit. We then correlate $\eta$ to the displacement $D$ of the center-most stripe with respect to the center position $x_0$.

\section{Time-of-flight experiments}
\begin{figure}
\includegraphics{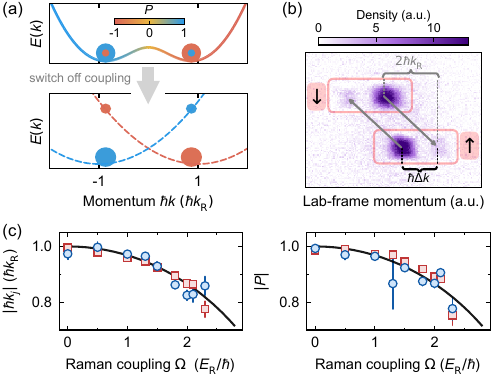}
\caption{\textbf{Momentum-space imaging scheme. } 
(a) Top panel: dispersion relation of the spin-orbit-coupled system with dressed BECs in both minima. The blue and red circles represent the left (right) dressed BECs, being composed of the majority blue (red) bare spin state and a smaller admixture of the red (blue) state. Bottom panel: projection onto the bare states when switching off the Raman coupling, yielding four clouds with different spin and momentum. 
(b) Example of a spin-resolved time-of-flight image. The momentum transfer $2\hbar k_\mathrm{R}$ is fixed by the Raman beams and independent of the spin-orbit coupling strength $\Omega$ (grey arrows), while the relative momentum $\hbar\Delta k$ between the dressed BECs is variable. It is the interference between same spin clouds separated by $\hbar\Delta k$ that yields the stripe spacing, which can thus also vary.
(c) Left panel: Momentum $|\hbar k_j|$ of the left $j=l$ (blue) and right $j=r$ (red) state versus $\Omega$ measured from the momentum of the clouds in time-of-flight. Right panel: Absolute value of the polarization $|P|$ of the left (blue) and right (red) Raman-dressed state vs. $\Omega$. Errorbars denote one eom. Black lines: single-particle predictions at the dispersion minima $k_{\ell}(k_r)$.}
\label{Fig:M4}
\end{figure}

\textit{Data collection.} For the time-of-flight measurements, we abruptly switch off the Raman coupling, which projects the dressed BECs back onto the bare states and leads to two distinct momentum components per spin state, see Figure \ref{Fig:M4}(a). We then let the atoms expand for $14.2$\,ms, during which we apply a magnetic field gradient in $z$ direction to perform Stern-Gerlach separation of the spin states. Finally, we image the atomic distribution along the $(x+y)$ axis and obtain spin resolved images of the momentum distribution, see example in Figure \ref{Fig:M4}(b). 

\textit{Data analysis.} The stripe spacing (grey squares in Figure~\ref{Fig:Fig2}(a)) is extracted from the relative momentum of the two central momentum clouds, and assuming a total Raman momentum transfer of $2\hbar k_\mathrm{R}$. We take data vs. holding time, such that we can infer mean values despite the presence of collective excitations. This technique thus allows to measure the stripe spacing for low spin-orbit coupling strength and very polarized mixtures. However, as opposed to the in situ data, it relies on assumptions regarding the spin-orbit coupling. For the frequency measurements of the stripe compression mode of Figure~\ref{Fig:Fig4}(b), we determine the momentum difference $\Delta k$ of the two minima in the frame co-moving with the Raman coupling fields from the lab-frame momentum of the two central peaks $(k_{\uparrow,0}^\mathrm{lab},k_{\downarrow,0}^\mathrm{lab})$. To this end, we use the relation $\Delta k=k_{\uparrow,0}^\mathrm{lab}-k_{\downarrow,0}^\mathrm{lab}+2k_\mathrm{R}$. For the dipole oscillation frequency, we measure the total momentum $k_\mathrm{tot} =k_{\uparrow,0}^\mathrm{lab}+k_{\downarrow,0}^\mathrm{lab}$.

\textit{Comparison to single-particle predictions.} To verify our system preparation and data collection procedures, we measure the group momentum and polarization of the spin-orbit-coupled BEC for different spin-orbit coupling strengths, see Figure \ref{Fig:M4}(c). The data points are consistent with the single-particle predictions for Raman-dressed states at the minima of their dispersion relation, $\left|k_j/k_R\right|=\sqrt{1-[\hbar \Omega/(4E_{\mathrm{R}})]^2}$ and $P=\tilde{\delta}(k_j)/\tilde{\Omega}$, with $j=\ell, r$.

\section{The mixture model}

\emph{Raman-coupled BECs.} 
We consider a BEC subjected to two Raman beams of frequency difference $\Delta\omega_\mathrm{R}$. They couple states $\ket{\downarrow}$ and $\ket{\uparrow}$, and impart a recoil momentum $\hbar k_{\mathrm{R}}$ per beam along the $x$ axis. In the laboratory frame, the gas is described by the spinor wavefunction $\{\ket{\uparrow,k^{\mathrm{lab}}},\ket{\downarrow,k^{\mathrm{lab}}}\}$. We perform a gauge transformation $\mathcal{U}=\exp\left[i\left(\Delta \omega_\mathrm{R} t - 2k_{\mathrm{R}}x\right){\sigma}_z/2\right]$ to the frame rotating at $\Delta\omega_\mathrm{R}$ and the atomic basis $\{\ket{\uparrow,k+k_{\mathrm{R}}},\ket{\downarrow,k-k_{\mathrm{R}}}\}$ \cite{Higbie2004, Spielman2009} which, unless indicated otherwise, is the one used throughout the paper. 

By diagonalizing the corresponding single-particle kinetic Hamiltonian $\mathcal{H}_{\mathrm{kin}}$, which corresponds to Eq. \eqref{H_SO} from the main text, we obtain two dressed bands 
\begin{equation}
\epsilon_\pm(\mathbf{k}) =\hbar^2(\mathbf{k}^2+k_\mathrm{R}^2)/(2m) \pm \hbar\tilde\Omega(k)/2
\end{equation}
and their corresponding higher and lower energy dressed states
\begin{align}
\ket{+, \mathbf{k}} &=  \mathcal{S}(k)\ket{\uparrow, \mathbf{k}} + \mathcal{C}(k)\ket{\downarrow, \mathbf{k}},\cr  
\ket{-, \mathbf{k}} &=  -\mathcal{C}(k)\ket{\uparrow, \mathbf{k}} + \mathcal{S}(k)\ket{\downarrow, \mathbf{k}},
\end{align}
with $\tilde\Omega=\sqrt{\Omega^2 + \tilde\delta^2}$, $\tilde\delta= \delta-2\hbar k_\mathrm{R}k/m$, $\mathcal{C}(k) = \sqrt{(1 + \tilde{\delta}/\tilde{\Omega})/2}$, and $\mathcal{S}(k) = \sqrt{(1 - \tilde{\delta}/\tilde{\Omega})/2}$. Here, $k = \mathbf{k}\cdot \mathbf{\hat{e}_x}$ is the momentum along the $x$ axis.

\emph{Lower energy band Hamiltonian.} 
To treat the many-body problem, we introduce the field operators $\hat{\phi}^{\dagger}_{\pm}(\mathbf{k})$ and $\hat{\phi}_{\pm}(\mathbf{k})$ that create and annihilate a particle in band state $\ket{\pm, \mathbf{k}}$, respectively. Moreover, we consider that the BEC does not occupy the higher dressed band. This assumption is valid for the preparation procedure employed in the experiment, for which the excitations remain small compared to the band separation $\tilde{\Omega}$. It allows us to truncate the description of the system to the lower dressed band $\ket{-,\mathbf{k}}$. Then, the many-body Hamiltonian reads 
\begin{equation}\label{Seq_int_Ham_lower_band}
\hat{H}\simeq \int \frac{\mathrm{d}^3 \mathbf{k}}{(2\pi)^3} \hat{\phi}_{-}^{\dagger} \epsilon_{-}\hat{\phi}_{-} + \frac{1}{2}\int \prod_{j=1}^4\frac{\mathrm{d}^3 \mathbf{k}_j}{(2\pi)^3} \hat{\mathcal{H}}_\mathrm{int},
\end{equation}
with 
\begin{align}
\hat{\mathcal{H}}_\mathrm{int} = &\hat{\phi}_{-}^{\dagger}(\mathbf{k}_4)\hat{\phi}_{-}^{\dagger}(\mathbf{k}_3)\hat{\phi}_{-}(\mathbf{k}_2)\hat{\phi}_{-}(\mathbf{k}_1) \chi(\mathbf{k}_1,\mathbf{k}_2,\mathbf{k}_3,\mathbf{k}_4) \cr
&\times \frac{4\pi \hbar^2}{m}\delta^{(3)}(\mathbf{k}_4+\mathbf{k}_3-\mathbf{k}_2-\mathbf{k}_1),
\end{align}
which describes two-body collisions between atoms of incoming momenta $\mathbf{k}_1$ and $\mathbf{k}_2$ and outgoing momenta $\mathbf{k}_3$ and $\mathbf{k}_4$. As pointed out in Ref. \cite{Williams_science_2012} and discussed in detail in our previous works \cite{Cabedo2021a,Cabedo2021b,Froelian2022, Chisholm2022}, due to their momentum-dependent spin composition the dressed atoms experience modified scattering amplitudes
\begin{align}
\chi(\mathbf{k}_1,\mathbf{k}_2,\mathbf{k}_3,\mathbf{k}_4) = &a_{\uparrow \uparrow} \mathcal{C}(k_{2}) \mathcal{C}(k_{4}) \mathcal{C}(k_{1}) \mathcal{C}(k_{3})\cr+&a_{\downarrow \downarrow} \mathcal{S}(k_{2}) \mathcal{S}(k_{4}) \mathcal{S}(k_{1}) \mathcal{S}(k_{3}) \cr+&a_{\uparrow \downarrow}\mathcal{C}(k_{2}) \mathcal{C}(k_{4}) \mathcal{S}(k_{1}) \mathcal{S}(k_{3})\cr+&a_{\uparrow \downarrow}\mathcal{S}(k_{2}) \mathcal{S}(k_{4}) \mathcal{C}(k_{1}) \mathcal{C}(k_{3}),
\end{align}
where $k_{j} = \mathbf{k}_j\cdot \mathbf{\hat{e}_x}$.

\emph{Presentation of the mixture model. } We consider the regime where the spin-orbit-coupled BEC has a single-particle dispersion relation $\epsilon_-$ with two minima along the $x$ direction, as required for supersolidity. It corresponds to the conditions $\hbar \Omega<4E_\mathrm{R}$ and $\hbar\vert\delta\vert <4E_\mathrm{R}\left(1-\left[\hbar \Omega /\left(4 E_\mathrm{R}\right)\right]^{2 / 3}\right)^{3 / 2}$. The positions of the minima for the left ($\ell$) and right ($r$) wells, which we denote $k_r > 0 $ and $k_\ell < 0$, fulfill the recursive equation $k_j / k_\mathrm{R}= -\tilde{\delta}\left(k_j\right) / \tilde{\Omega}\left(k_j\right)$, with $j=\ell,r$. Because the gas has a small momentum spread around such wells, we can describe its low energy properties by truncating the lower dressed band Hamiltonian Eq. \eqref{Seq_int_Ham_lower_band} around them. Formally, we recast the momentum around $\ell$ ($r$) as $\mathbf{q}=(k-k_{\ell(r)},k_y,k_z)$ and replace the field operator of the dressed band $\hat{\phi}_-(\mathbf{q})$ by two new field operators, one per well $\hat{\varphi}_{\ell(r)}(\mathbf{q})=\hat{\phi}_-(k_{\ell(r)}\mathbf{e_x}+\mathbf{q})$ \cite{Higbie2004, Chisholm2023}. The corresponding inverse-Fourier-transformed operators are $\hat\varphi_{\ell(r)}(\mathbf{r})$, but the physically-relevant operators in position space are instead $\hat\varphi_{\ell (r)}^\prime(\mathbf{r}) = \exp(-i k_{\ell(r)}x)\hat\varphi_{\ell(r)}(\mathbf{r})$, which take into account the phase difference between the two fields. We name this effective field theory the mixture model. It allows us to treat the spin-orbit-coupled BEC as an effective mixture of condensates with momenta centered at $k_{\ell(r)}$ and modified kinetic and interaction properties, which depend on the spin-orbit coupling parameters. While Ref. \cite{Lin2011} exploited it to describe the many-body phase diagram of a spin-orbit-coupled BEC treating $\Omega$ as a perturbation, in this work we show that it yields accurate analytical expressions for the relevant physical observables beyond the perturbative regime, provides an intuitive understanding of the behavior of the system, and emphasizes how supersolidity is straightforward to understand in the spin-orbit-coupled platform. We use this model to produce all theoretical predictions presented in the main text. 

\emph{Derivation of the mixture model Hamiltonian. } In the mixture model, the kinetic energy term of the Hamiltonian is simplified by performing a second-order Taylor expansion of the dispersion relation in $\mathbf{q}$, which yields $\varepsilon_{-} \approx \hbar^2\left(k_{\ell(r)}^2 + \mathbf{k}_\perp^2\right)/(2m) - E_\mathrm{R}  + \hbar^2 (k-k_{\ell(r)})^2/(2 m_{\ell(r)}^*) + \hbar\delta k_\mathrm{R}/(2k_{\ell (r)}) $. The result corresponds to that of conventional BECs, but with an effective mass $m_j^*$ along the $x$ direction that encapsulates the modified curvature of the dressed-band dispersion at the minima $j=\ell,r$
\begin{equation}\label{Seq_effective_mass}
\frac{1}{m_j^*}= \frac{1}{m}\left[1-4 \frac{k_j / k_{\mathrm{R}}}{4 k_j / k_{\mathrm{R}}-\hbar \delta / E_\mathrm{R}}\left(1-k_j^2 / k_{\mathrm{R}}^2\right)  \right].  
\end{equation}
Such effective mass is crucial to understand the dipole excitations of the system measured in Figures \ref{Fig:Fig3}(b) and \ref{Fig:Fig4}(b).

To be consistent with the kinetic energy expansion to second order in momentum around $\mathbf{q}$, we expand the interacting part of the Hamiltonian to zero order in $\mathbf{q}$. Due to momentum conservation, the only allowed collisions are of the form, $i\, i \to i\, i$ (intrawell) or $i\,j \to j\, i $ (interwell), and thus
\begin{align}
\hat{\mathcal{H}}_\mathrm{int} \approx \sum_{i= \ell, r}\sum_{j= \ell, r}&  \frac{4\pi \hbar^2 a_{ij}}{m}  \hat{\varphi}_{i}^{\dagger}(\mathbf{q}_4)\hat{\varphi}_{j}^{\dagger}(\mathbf{q}_3)\hat{\varphi}_{i}(\mathbf{q}_2)\hat{\varphi}_{j}(\mathbf{q}_1)  \cr
&\times\delta^{(3)}(\mathbf{q}_4+\mathbf{q}_3-\mathbf{q}_2-\mathbf{q}_1),
\end{align}
where $a_{ij} = \chi(k_i,k_j,k_i,k_j)$ are the effective scattering lengths of the mixture.
Explicitly, the effective scattering lengths take the form 
\begin{align}\label{Seq_aij}
a_{jj} = &a_{\uparrow\uparrow}\frac {(1-k_j/k_{\mathrm{R}})^2}4 + a_{\downarrow\downarrow}\frac {(1+k_j/k_{\mathrm{R}})^2}4 +
a_{\uparrow\downarrow} \frac{(1-k_j^2/k_{\mathrm{R}}^2)}2 ,\cr
a_{\ell r} = 
&a_{\downarrow\downarrow}\frac {(1+k_{\ell}/k_{\mathrm{R}})(1+k_r/k_{\mathrm{R}})}2 \cr + 
&a_{\uparrow\uparrow}\frac {(1-k_{\ell}/k_{\mathrm{R}})(1-k_r/k_{\mathrm{R}})}2 \cr + 
&a_{\uparrow\downarrow}\frac{1 - k_{\ell} k_r/k_{\mathrm{R}}^2 + \sqrt{1-k_{\ell}^2/k_{\mathrm{R}}^2}\sqrt{1-k_r^2/k_{\mathrm{R}}^2}}{2},
\end{align}
where we have used the location of the minima $k_j / k_\mathrm{R}= -\tilde{\delta}\left(k_j\right) / \tilde{\Omega}\left(k_j\right)$ and the expressions for $\mathcal{C}$ and $\mathcal{S}$. The dependence of the $a_{ij}$ with the spin-orbit coupling strength are used in Figures \ref{Fig:Fig1}(c), (d) and \ref{Fig:Fig4}(b), (c) of the main text. 

In position space, the complete Hamiltonian of the mixture model takes the form 
\begin{align}
\hat{H}&\approx \sum_{j={\ell,r}}\int \mathrm{d}^3\mathbf{r}\,\hat\varphi_j^{\prime \dagger}\Big(-\frac{\hbar^2 }{2 m_j^*}\frac{\partial^2}{\partial x^2}-\frac{\hbar^2 }{2 m}\nabla^2_\perp +\frac{\Delta_j}2\Big) \hat{\varphi}_j^\prime\cr
&+
\frac{1}{2}\sum_{ij = \ell,r} g_{ij}\int \mathrm{d}^3\mathbf{r} \hat\varphi_i^{\prime \dagger}(\mathbf{r})\hat\varphi_j^{\prime \dagger}(\mathbf{r})\hat{\varphi}_i^\prime(\mathbf{r})\hat{\varphi}_j^\prime(\mathbf{r}),
\end{align}
where we have defined $g_{ij} = 4\pi \hbar^2 a_{ij}/m$ and introduced the energy difference between the left and the right wells $\Delta_{\ell, r} = \pm (\epsilon_-(k_\ell) - \epsilon_-(k_r))$.

\emph{Supersolid stripe phase. } 
Within the mixture model, supersolidity occurs when dressed-state interactions favor both the spatial overlap of the two modes $\hat{\varphi}'_r$ and $\hat{\varphi}'_\ell$, i.e., the gas is in the miscible regime of the dressed mixture, and their simultaneous macroscopic occupation. 

Writing their mean-field wavefunctions as $\varphi_{j}^\prime(\mathbf{r}) = \vert \varphi_{j}^\prime(\mathbf{r}) \vert \exp(i\theta_{s,j})$, with $j=\ell,r$, the atomic wavefunctions in the bare basis $\uparrow,\downarrow$ take the form
$\phi_\downarrow = \sum_{j=\ell,r}
\mathcal S(k_j)\vert \varphi_j\vert  \exp(i\theta_{s,j}-ik_j x)$ and $
\phi_\uparrow = \sum_{j=\ell,r}
\mathcal C(k_j)\vert \varphi_j\vert  \exp(i\theta_{s,j}-ik_j x)$,
from which we compute the total mean-field density 
\begin{align}\label{Seq_meanfield_dens}
n(\mathbf{r})&=\left|\phi_{\downarrow}(\mathbf{r})\right|^2+\left|\phi_{\uparrow}(\mathbf{r})\right|^2 =\left|\varphi_\ell^\prime(\mathbf{r})\right|^2+\left|\varphi_r^\prime(\mathbf{r})\right|^2 \cr
&+2\left|\varphi_\ell^\prime(\mathbf{r})\right|\left|\varphi_r^\prime(\mathbf{r})\right|\Big[\mathcal{C}(k_r)\mathcal{C}(k_\ell) \cr
&\,\,\,+\mathcal{S}(k_r)\mathcal{S}(k_\ell)\Big] \cos \left[\left(k_r-k_\ell\right) x+\theta_s\right], 
\end{align}
where $\theta_s = \theta_{s,\ell}-\theta_{s,r}$ is the relative phase between the $\ell$ and $r$ condensates. For $\left|\varphi_\ell^\prime(\mathbf{r})\right|\left|\varphi_r^\prime(\mathbf{r})\right| > 0$ and $\Omega > 0$, the density profile of the spin-orbit-coupled BEC exhibits a spatial modulation along the $x$ direction. Its period is given by the inverse of the momentum difference $\Delta k=k_r-k_{\ell}$ between the two dispersion minima, as observed in Figure \ref{Fig:Fig2}(b). 

We determine the boundaries of the supersolid stripe phase from the mean-field energy density of the dressed mixture in the homogeneous limit
\begin{align}\label{Seq_energy_functional_hom}
\mathcal{E} &= \frac 12 \left[ g_{rr} \,n_  r^2 + g_{\ell\ell} \,n_\ell^2 + 2 g_{\ell r}  \,n_\ell n_r + \Delta_r (n_r-n_\ell) \right]\cr
&= \frac 12 \bar n^2\left[ \tilde g_{nn} + \tilde g_{ss} \left(S - S_0 \right)^2 - \tilde g_{ss}  S_0^2\right].
\end{align}
Here we have introduced the polarization of the mixture $S = \int \mathrm{d}^3\mathbf{r} ( \vert \varphi_l \vert^2 - \vert \varphi_r \vert^2 )/N$, and written the densities of the effective mixture components as $n_{\ell, r} = \bar n(1\pm S)/2 $, where $\bar n$ is the mean density of the gas. Moreover, we have defined $S_0 = -[\tilde g_{ns} - \Delta_r/(2\bar n)]/\tilde g_{ss}$, where $\tilde g_{nn} = (g_{\ell \ell}+g_{r r}+2g_{\ell r})/4$, $\tilde g_{ss} = (g_{\ell \ell}+g_{r r}-2g_{\ell r})/4$ and $\tilde g_{ns} = (g_{\ell\ell}-g_{rr})/4$. To our order of approximation, the effective Hamiltonian does not include any terms that allow population transfer between the two dressed modes. While spatial confinement and thermal effects enable the exchange, previous experiments on spin-orbit-coupled BECs have shown that their rates are not significant on typical experimental timescales \cite{Lin2011}. Therefore, we can treat the effective polarization $S$ as a conserved quantity. It relates to the bare basis spin polarization $P = \int \mathrm{d}^3\mathbf{r} ( \vert \phi_\uparrow \vert^2 - \vert \phi_\downarrow \vert^2 )/N=(N_{\uparrow}-N_{\downarrow})/N$ simply by $P=-(1-S)k_r/(2k_\mathrm{R})-(1+S)k_{\ell}/(2k_\mathrm{R})$.

From Eq. \eqref{Seq_energy_functional_hom}, it follows that the effective mixture is miscible when $\tilde g_{ss} > 0$, and that both minima are occupied in the ground state when $\vert S \vert < 1$. The ground state polarization is then given by $S = S_0$. The boundaries of the stripe phase are therefore found by imposing $\vert S_0  \vert = 1$ and $\tilde g_{ss} \geq0$, which gives the condition for the critical values of $\Omega$ and $\delta$
\begin{equation}\label{Seq_boundaries_stripe}
\Delta_{r}(\Omega_\mathrm{c}, \delta_\mathrm{c}) = 2\tilde g_{ns}(\Omega_\mathrm{c}, \delta_\mathrm{c}) \bar n \pm 2\tilde g_{ss}(\Omega_\mathrm{c}, \delta_\mathrm{c}) \bar n,
\end{equation}
used to determine the boundaries of the stripe-to-plane-wave phase transition in the phase diagram of Figure \ref{Fig:Fig1}(d). Along $\delta=0$, and for $g_{jj} > g_{ii}$, Eq. \eqref{Seq_boundaries_stripe} implies $a_{ii}(\Omega_\mathrm{c}) = a_{\ell r }(\Omega_\mathrm{c})$, as illustrated in Figure \ref{Fig:Fig1}(c). In these conditions, and within our effective model, $\Omega_\mathrm{c}$ becomes independent of the atomic density. Likewise, the contrast of the stripes $C$ becomes density independent. Indeed, by using the expressions for the band minima at $\delta=0$, $k_{\ell, r} = \mp k_\mathrm{R} \sqrt{1-\left[\hbar \Omega /\left(4 E_\mathrm{R}\right)\right]^2}$, we retrieve the contrast from the density profile of Eq. \eqref{Seq_meanfield_dens}, which reads $C = \sqrt{1 - (\tilde{g}_{ns}/\tilde{g}_{ss})^2} \hbar \Omega/(4 E_\mathrm{R})$. 
As discussed in the main text, it increases linearly with $\Omega$.

\section{Collective mode frequencies}

\emph{Ehrenfest model.}
The mixture model can also be employed to derive analytical expressions for the collective excitations of the spin-orbit-coupled BEC in a harmonic trap investigated in Figures \ref{Fig:Fig3}(b) and \ref{Fig:Fig4}(a)--(c). Here we restrict our analysis to the conditions of the experiment: $g_{rr}, g_{\ell\ell} > g_{\ell r}$ so that the ground state of the mixture is not fully magnetized, a spin-independent harmonic trapping potential $V_{\mathrm{ho}}$, and a Raman detuning $\delta = 0$, for which $\Delta_{\ell,r} = 0$ and the effective masses of both dressed components are identical $1/m^*_{\ell,r} = 1/m^* = (1-\left[\Omega/(4E_\mathrm{R})\right]^2)/m$.    

The dipole modes of a mixture along the $x$ direction involve in-phase and out-of-phase oscillations of the two components and can be characterized by the oscillations of the first moments of the distribution $\delta x_j \equiv \langle x\rangle_j = \int x n_j \mathrm{d}^3\mathbf{r} $, where $n_j = \varphi_j^*\varphi_j /N_j$ is the density profile of the $j$ component normalized to $1$ and $N_j$ is its atom number. Following the work of Refs. \cite{Fort2022, Cavicchioli2022} on conventional mixtures, we characterize such oscillations using the Ehrenfest theorem, which states that the time evolution of expectation values follows classical laws of motion 
\begin{equation}\label{Seq_Ehrenfest_eq}
\frac{\partial^2}{\partial t^2}\delta x_{\ell(r)} = -\frac{1}{m^*} \langle \partial_x V_{\ell(r)} (\vec{r},t) \rangle.
\end{equation}
Here we have introduced the total mean-field potential acting on the $j$ component
\begin{equation}\label{Seq_Vtot}
V_j = V_\mathrm{ho}(\vec{r}) +  g_{jj}N_jn_j(\vec{r},t) +g_{ji}N_in_i(\vec{r},t),
\end{equation}
with $i\neq j$. Combining the two expressions yields
\begin{equation}\label{Seq_Ehrenfest_eq2}
\frac{\partial^2}{\partial t^2}\delta x_j = \bar\omega_{x}^2  \delta x_j - \frac{g_{\ell r} N_i}{m^*}\int n_j \partial_x n_i \mathrm{d}^3\mathbf{r},
\end{equation}
where $\bar\omega_{x} \equiv \omega_{x}\sqrt{m/m^*}$. In the following, we consider the regime of small-amplitude rigid oscillations, and assume that $n_j(\vec{r},0)$ are even functions. Then, $n_j(\vec{r},t) \simeq n_j(\vec{r},0) - \partial_x n_j(\vec{r},0)\delta x_j$ and the integral reads $\int n_j \partial_x n_i \mathrm{d}^3\mathbf{r} = -(\delta x_j-\delta x_i) I$, where we have defined $I \equiv \int \partial_x n_j(\vec{r},0)\partial_x n_i(\vec{r},0) \mathrm{d}^3\mathbf{r}$.

Equations \eqref{Seq_Ehrenfest_eq2} can be rewritten in matrix form 
\begin{align}\label{Seq_diff_eq_dipole_modes}
\frac{\partial^2}{\partial t^2}\left(\begin{array}{c} \delta x_r \\ \delta x_\ell\end{array}\right)&\approx-\left(\begin{array}{cc} \bar \omega_x^2-\xi_r & \xi_r \\ \xi_\ell & \bar \omega_x^2-\xi_\ell\end{array}\right)\left(\begin{array}{c} \delta x_r \\ \delta x_\ell\end{array}\right),
\end{align}
where we have defined the coefficients $\xi_j \equiv g_{\ell r} N_i I/m^*$. The two dipole mode frequencies are given by the eigenfrequencies of the normal modes $\omega_\mathrm{D}^2$ and $\omega_\mathrm{SC}^2$. 

The first eigenvalue corresponds to the in-phase dipole mode $\omega_\mathrm{D}$, which oscillates at the effective trap frequency $\omega_\mathrm{D}= \bar\omega_{x}$. It differs from the trap frequency $\omega_x$ by a factor $\sqrt{m/m^*}$, corresponding to the effective mass correction along the $x$ direction. Its dependency on the spin-orbit coupling strength $\Omega$ is clearly visible in Figure \ref{Fig:Fig4}(b). The correct value of $\omega_{\mathrm{D}}$ must also be taken into account to interpret the breathing oscillation experiments of Figure \ref{Fig:Fig3}(b).

The second eigenvalue corresponds to the out-of-phase spin-dipole mode. In the mixture model, it corresponds to an oscillatory displacement of the $\varphi'_{\ell}$ and $\varphi'_r$ condensates in opposite directions, leading to a periodic variation of their relative momentum. Since the latter determines the stripe spacing in position space, see Eq. \eqref{Seq_meanfield_dens}, the out-of-phase spin-dipole mode corresponds to an oscillation of the stripe spacing, i.e., to the stripe compression mode. Its frequency $\omega_\mathrm{SC} = \sqrt{\bar\omega_{x}^2 - \xi_r - \xi_\ell}$ depends on the interactions of the system and exhibits the softening at the supersolid stripe-to-plane-wave transition investigated in Figures \ref{Fig:Fig4}(b) and (c).

A key insight provided by our mixture model is that, within its regime of validity, the in-phase and out-of-phase modes are orthogonal (density- and spin-like, respectively) and remain decoupled.  As a result, density-like modes such as the dipole and breathing modes, which preserve the relative momentum between $\varphi'_\ell$ and $\varphi'_r$, do not couple to spin-like modes such as the stripe compression mode. The situation is very different in dipolar supersolids, where density and crystal excitations generally hybridize \cite{TanziNature2019,Natale2019, Hertkorn2019}.

\emph{Analytical expression for $\omega_{\mathrm{SC}}$. }
We derive an analytical expression for the stripe compression mode in the Thomas-Fermi regime, where the thermodynamic limit of the trapped system and the phase transition are well defined. To this end, we extremize the Thomas-Fermi energy functional
\begin{align}\label{Seq_mean_field_energy}
E[n_j] \approx \int \mathrm{d}^3\mathbf{r} \Bigg[ &V_\mathrm{ho} (N_r n_r + N_\ell n_\ell ) + \frac{1}{2}(g_{rr} N_r^2 n_r^2 \cr & + g_{\ell\ell} N_\ell^2 n_\ell^2 +2g_{\ell r} N_\ell N_r n_\ell n_r) \Bigg].
\end{align}
under the constraints $\int n_j \mathrm{d}^3\mathbf{r}  = 1$. From the result, we obtain the explicit density profiles of the two components $n_j (\mathbf{r},0) = \max \left[ \alpha_j - \beta_j(\omega_x^2 x^2+\omega_y^2 y^2+\omega_z^2 z^2)\,, \,\, 0 \right]$, where $\beta_j = m(g_{ii}-g_{\ell r})/[2 N_j(g_{\ell\ell}g_{rr}-g_{\ell r}^2)]$ and $\alpha_j = \left[15 \omega_x \omega_y \omega_z/(8 \pi)\right]^{2 / 5} \beta^{3 / 5}_j$, and a general expression for the stripe compression mode frequency
\begin{equation}\label{Seq_omega_sd_betaj}
\omega_\mathrm{SC}^2 =  \bar\omega_{x}^2 - \xi_r - \xi_\ell = \bar\omega_{x}^2 - \frac{2g_{\ell r} N \bar\omega_{x}^2}{m} \min\left( \beta_r, \beta_\ell \right).
\end{equation}
Given that $\beta_\ell$ ($\beta_r$) is a monotonously decreasing (increasing) function of the polarization of the mixture $S = (N_\ell - N_r)/(N_\ell + N_r)$, $\omega_\mathrm{SC}^2(S)$ will be minimal at the crossing $\beta_\ell(S_\mathrm{min}) = \beta_r(S_\mathrm{min})$, which implies
\begin{equation}\label{Seq_Pmin}
S_\mathrm{min} = \frac{g_{rr}-g_{\ell \ell}}{g_{\ell\ell}+g_{rr}-2g_{\ell r}}.
\end{equation}
Remarkably, this same condition on the polarization minimizes the mean-field energy of the mixture Eq. \eqref{Seq_mean_field_energy}. Therefore, the ground state of the spin-orbit-coupled system at $\delta = 0$ and in the Thomas-Fermi regime fulfills
\begin{align}\label{Seq_omega_sd_gs}
\omega_\mathrm{SC}^0 &= \mathrm{min}(\omega_\mathrm{SC}) = \omega_x \sqrt{\frac{m}{m^*}} \sqrt{1 - \frac{g_{\ell r} (g_{\ell \ell}+g_{rr}-2g_{\ell r})}   {g_{\ell\ell}g_{rr}- g_{\ell r}^2}},
\end{align}
which can be explicitly written in terms of the spin-orbit coupling strength $\Omega$ by using the expressions of the effective scattering lengths, the position of the minima and the effective scattering length. This result is independent of the total number of particles, as expected in the thermodynamic limit. It yields the theoretical predictions for the stripe compression mode frequency included in Figures \ref{Fig:Fig4}(b), (c) of the main text.

\emph{Stripe compression mode softening at the supersolid phase transition. }
Our analytical expression for the stripe compression mode frequency Eq. \eqref{Seq_omega_sd_gs} predicts a softening of the mode at the phase transition between the supersolid stripe phase and the plane-wave phase, where $g_{\ell r} \rightarrow \min(g_{\ell\ell},g_{rr})$ and thus $\omega_\mathrm{SC}^{0} \rightarrow 0$. In the thermodynamic limit, this allows one to locate the phase transition point $\Omega_\mathrm{c}$ by experimentally finding for each value of the Raman coupling $\Omega$ the polarization $P$ that yields the minimal stripe compression mode frequency, which is also that of the ground state. Because minimizing this mode frequency allows us to set the metastable states apart from the ground state, it is a robust observable to locate the phase transition. Indeed, these metastable states continue to exist in the non-supersolid plane-wave phase and exhibit lifetimes that can be larger than typical experimental timescales \cite{Lin2011}. Therefore, they prevent the contrast of the stripes from being a good experimental observable to locate the phase transition, and they are the reason why the stripe compression mode softening is preferred.

\begin{figure}
\includegraphics{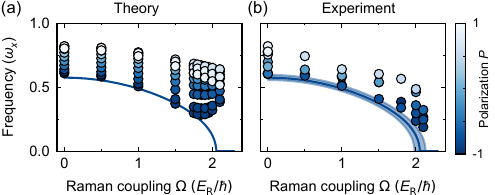}
\caption{\textbf{Effect of polarization on the stripe compression mode frequency.}
(a) Stripe compression mode frequency $\omega_{\mathrm{SC}}$ vs. Raman coupling $\Omega$ ($x$ axis) and polarization $P$ (colorscale) for a finite size system (circles) compared to the thermodynamic limit prediction (solid line), as computed from the Ehrenfest model and the density profiles obtained from Gross-Pitaevskii numerical simulations. At each value of $\Omega$, the polarization that minimizes $\omega_{\mathrm{SC}}$ converges to that of the ground state in the thermodynamic limit, except very close to the transition point where a finite gap and a small shift in $\Omega_{\mathrm{c}}$ take place. 
(b) Experimental measurements, which are in good agreement with the theory predictions except in the close vicinity of the transition point (see text). The shaded area indicates the magnetic field uncertainty.}
\label{Fig:M5}
\end{figure}

In any experiment, however, finite size effects will appear. To estimate their role for our experimental parameters, we numerically solve the Gross-Pitaevskii equations of the dressed mixture as a function of $\Omega$ for different values of $P$, and extract the corresponding value of $\omega_\mathrm{SC}$ by computing the integral $I$ from the obtained profiles. Figure \ref{Fig:M5}(a) shows the numerical values of $\omega_\mathrm{SC}$ obtained by this procedure as a function of $\Omega$ and $P$. The parameters are chosen to match the experimental conditions of the time-of-flight measurements of Figure \ref{Fig:Fig4}(b), see table \ref{tab:exp_params}. Away from the transition, we find $\mathrm{min}(\omega_\mathrm{SC})$ to be well approximated by the Thomas-Fermi prediction Eq. \eqref{Seq_omega_sd_gs}. In the vicinity of the transition, however, the numerical results show a finite gap, and a shift of the position of the global minimum towards values of $\Omega$ below $\Omega_{\mathrm{c}}$. 

These predictions are compared to our experimental results in Figure\,\ref{Fig:M5}(b). There, for each value of $\Omega$, we vary the polarization $P$ of the spin-orbit-coupled BEC by adjusting the length of the rf pulse before switching on the Raman coupling. The experimental data reproduces well the expected finite size gap, but lies closer to the thermodynamic limit predictions than to our finite-size numerical simulations near the phase transition point. This discrepancy is not completely explained by considering our magnetic field uncertainty (blue shaded area in Figure \ref{Fig:M5}(b)) which, however, does not take into account systematic errors in our scattering length model that could affect the resonance width \cite{Tanzi2018}. Another important experimental point is that, as we approach the phase transition, the absolute value of the polarization that minimizes the stripe compression mode frequency $|P_\mathrm{min}|$ increases. Our resolution of the phase transition point is thus also limited by our ability to prepare and detect very spin imbalanced clouds. Moreover, for the slowest oscillations close to $\Omega_\mathrm{c}$, we are affected by the limited lifetime of the gas due to inelastic photon scattering from the Raman beams. Finally, the numerical calculation of the density profiles might not capture all the excitation properties, and limitations in the Ehrenfest approach could also partly explain the discrepancy.


\begin{thebibliography}{60}%
\makeatletter
\providecommand \@ifxundefined [1]{%
 \@ifx{#1\undefined}
}%
\providecommand \@ifnum [1]{%
 \ifnum #1\expandafter \@firstoftwo
 \else \expandafter \@secondoftwo
 \fi
}%
\providecommand \@ifx [1]{%
 \ifx #1\expandafter \@firstoftwo
 \else \expandafter \@secondoftwo
 \fi
}%
\providecommand \natexlab [1]{#1}%
\providecommand \enquote  [1]{``#1''}%
\providecommand \bibnamefont  [1]{#1}%
\providecommand \bibfnamefont [1]{#1}%
\providecommand \citenamefont [1]{#1}%
\providecommand \@href[1]{\@@startlink{#1}\@@href}%
\providecommand \@@href[1]{\endgroup#1\@@endlink}%
\providecommand \@sanitize@url [0]{\catcode `\\12\catcode `\$12\catcode
  `\&12\catcode `\#12\catcode `\^12\catcode `\_12\catcode `\%12\relax}%
\providecommand \@@startlink[1]{}%
\providecommand \@@endlink[0]{}%
\providecommand \@url [1]{\endgroup\@href {#1}{\urlprefix }}%
\providecommand \urlprefix  [0]{URL }%
\providecommand \doibase [0]{https://doi.org/}%
\providecommand \selectlanguage [0]{\@gobble}%
\providecommand \bibinfo  [0]{\@secondoftwo}%
\providecommand \bibfield  [0]{\@secondoftwo}%
\providecommand \translation [1]{[#1]}%
\providecommand \BibitemOpen [0]{}%
\providecommand \bibitemStop [0]{}%
\providecommand \bibitemNoStop [0]{.\EOS\space}%
\providecommand \EOS [0]{\spacefactor3000\relax}%
\providecommand \BibitemShut  [1]{\csname bibitem#1\endcsname}%
\let\auto@bib@innerbib\@empty

\bibitem [{\citenamefont {Boninsegni}\ and\ \citenamefont {Prokof'ev}(2012)}]{Boninsegni2012}%
  \BibitemOpen
  \bibfield  {author} {\bibinfo {author} {\bibfnamefont {M.}~\bibnamefont {Boninsegni}}\ and\ \bibinfo {author} {\bibfnamefont {N.~V.}\ \bibnamefont {Prokof'ev}},\ }\bibfield  {title} {\bibinfo {title} {{Colloquium: Supersolids: What and where are they?}},\ }\href {https://doi.org/10.1103/RevModPhys.84.759} {\bibfield  {journal} {\bibinfo  {journal} {Rev. Mod. Phys.}\ }\textbf {\bibinfo {volume} {84}},\ \bibinfo {pages} {759} (\bibinfo {year} {2012})}\BibitemShut {NoStop}%
\bibitem [{\citenamefont {Recati}\ and\ \citenamefont {Stringari}(2023)}]{Recati2023}%
  \BibitemOpen
  \bibfield  {author} {\bibinfo {author} {\bibfnamefont {A.}~\bibnamefont {Recati}}\ and\ \bibinfo {author} {\bibfnamefont {S.}~\bibnamefont {Stringari}},\ }\bibfield  {title} {\bibinfo {title} {Supersolidity in ultracold dipolar gases},\ }\href {https://doi.org/10.1038/s42254-023-00648-2} {\bibfield  {journal} {\bibinfo  {journal} {Nat. Rev. Phys.}\ }\textbf {\bibinfo {volume} {5}},\ \bibinfo {pages} {735} (\bibinfo {year} {2023})}\BibitemShut {NoStop}%
\bibitem [{\citenamefont {Andreev}\ and\ \citenamefont {Lifshitz}(1969)}]{Andreev1969}%
  \BibitemOpen
  \bibfield  {author} {\bibinfo {author} {\bibfnamefont {A.~F.}\ \bibnamefont {Andreev}}\ and\ \bibinfo {author} {\bibfnamefont {I.}~\bibnamefont {Lifshitz}},\ }\bibfield  {title} {\bibinfo {title} {Quantum theory of defects in crystals},\ }\href {https://doi.org/10.1070/PU1971v013n05ABEH004235} {\bibfield  {journal} {\bibinfo  {journal} {Sov. Phys. JETP}\ }\textbf {\bibinfo {volume} {29}},\ \bibinfo {pages} {1107} (\bibinfo {year} {1969})}\BibitemShut {NoStop}%
\bibitem [{\citenamefont {Chester}(1970)}]{Chester1970}%
  \BibitemOpen
  \bibfield  {author} {\bibinfo {author} {\bibfnamefont {G.~V.}\ \bibnamefont {Chester}},\ }\bibfield  {title} {\bibinfo {title} {{Speculations on Bose-Einstein condensation and quantum crystals}},\ }\href {https://doi.org/10.1103/PhysRevA.2.256} {\bibfield  {journal} {\bibinfo  {journal} {Phys. Rev. A}\ }\textbf {\bibinfo {volume} {2}},\ \bibinfo {pages} {256} (\bibinfo {year} {1970})}\BibitemShut {NoStop}%
\bibitem [{\citenamefont {Leggett}(1970)}]{Leggett1970}%
  \BibitemOpen
  \bibfield  {author} {\bibinfo {author} {\bibfnamefont {A.~J.}\ \bibnamefont {Leggett}},\ }\bibfield  {title} {\bibinfo {title} {{Can a solid be "Superfluid"?}},\ }\href {https://doi.org/10.1103/PhysRevLett.25.1543} {\bibfield  {journal} {\bibinfo  {journal} {Phys. Rev. Lett.}\ }\textbf {\bibinfo {volume} {25}},\ \bibinfo {pages} {1543} (\bibinfo {year} {1970})}\BibitemShut {NoStop}%
\bibitem [{\citenamefont {L{\'{e}}onard}\ \emph {et~al.}(2017)\citenamefont {L{\'{e}}onard}, \citenamefont {Morales}, \citenamefont {Zupancic}, \citenamefont {Esslinger},\ and\ \citenamefont {Donner}}]{Leonard2017}%
  \BibitemOpen
  \bibfield  {author} {\bibinfo {author} {\bibfnamefont {J.}~\bibnamefont {L{\'{e}}onard}}, \bibinfo {author} {\bibfnamefont {A.}~\bibnamefont {Morales}}, \bibinfo {author} {\bibfnamefont {P.}~\bibnamefont {Zupancic}}, \bibinfo {author} {\bibfnamefont {T.}~\bibnamefont {Esslinger}},\ and\ \bibinfo {author} {\bibfnamefont {T.}~\bibnamefont {Donner}},\ }\bibfield  {title} {\bibinfo {title} {{Supersolid formation in a quantum gas breaking a continuous translational symmetry}},\ }\href {https://doi.org/10.1038/nature21067} {\bibfield  {journal} {\bibinfo  {journal} {Nature}\ }\textbf {\bibinfo {volume} {543}},\ \bibinfo {pages} {87} (\bibinfo {year} {2017})}\BibitemShut {NoStop}%
\bibitem [{\citenamefont {Li}\ \emph {et~al.}(2017)\citenamefont {Li}, \citenamefont {Lee}, \citenamefont {Huang}, \citenamefont {Burchesky}, \citenamefont {Shteynas}, \citenamefont {Topi}, \citenamefont {Jamison},\ and\ \citenamefont {Ketterle}}]{Li2017}%
  \BibitemOpen
  \bibfield  {author} {\bibinfo {author} {\bibfnamefont {J.~R.}\ \bibnamefont {Li}}, \bibinfo {author} {\bibfnamefont {J.}~\bibnamefont {Lee}}, \bibinfo {author} {\bibfnamefont {W.}~\bibnamefont {Huang}}, \bibinfo {author} {\bibfnamefont {S.}~\bibnamefont {Burchesky}}, \bibinfo {author} {\bibfnamefont {B.}~\bibnamefont {Shteynas}}, \bibinfo {author} {\bibfnamefont {F.~{\c{C}}.}\ \bibnamefont {Topi}}, \bibinfo {author} {\bibfnamefont {A.~O.}\ \bibnamefont {Jamison}},\ and\ \bibinfo {author} {\bibfnamefont {W.}~\bibnamefont {Ketterle}},\ }\bibfield  {title} {\bibinfo {title} {{A stripe phase with supersolid properties in spin-orbit-coupled Bose-Einstein condensates}},\ }\href {https://doi.org/10.1038/nature21431} {\bibfield  {journal} {\bibinfo  {journal} {Nature}\ }\textbf {\bibinfo {volume} {543}},\ \bibinfo {pages} {91} (\bibinfo {year} {2017})}\BibitemShut {NoStop}%
\bibitem [{\citenamefont {Tanzi}\ \emph {et~al.}(2019)\citenamefont {Tanzi}, \citenamefont {Lucioni}, \citenamefont {Fam\`a}, \citenamefont {Catani}, \citenamefont {Fioretti}, \citenamefont {Gabbanini}, \citenamefont {Bisset}, \citenamefont {Santos},\ and\ \citenamefont {Modugno}}]{Tanzi2019}%
  \BibitemOpen
  \bibfield  {author} {\bibinfo {author} {\bibfnamefont {L.}~\bibnamefont {Tanzi}}, \bibinfo {author} {\bibfnamefont {E.}~\bibnamefont {Lucioni}}, \bibinfo {author} {\bibfnamefont {F.}~\bibnamefont {Fam\`a}}, \bibinfo {author} {\bibfnamefont {J.}~\bibnamefont {Catani}}, \bibinfo {author} {\bibfnamefont {A.}~\bibnamefont {Fioretti}}, \bibinfo {author} {\bibfnamefont {C.}~\bibnamefont {Gabbanini}}, \bibinfo {author} {\bibfnamefont {R.~N.}\ \bibnamefont {Bisset}}, \bibinfo {author} {\bibfnamefont {L.}~\bibnamefont {Santos}},\ and\ \bibinfo {author} {\bibfnamefont {G.}~\bibnamefont {Modugno}},\ }\bibfield  {title} {\bibinfo {title} {Observation of a dipolar quantum gas with metastable supersolid properties},\ }\href {https://doi.org/10.1103/PhysRevLett.122.130405} {\bibfield  {journal} {\bibinfo  {journal} {Phys. Rev. Lett.}\ }\textbf {\bibinfo {volume} {122}},\ \bibinfo {pages} {130405} (\bibinfo {year} {2019})}\BibitemShut {NoStop}%
\bibitem [{\citenamefont {B\"ottcher}\ \emph {et~al.}(2019)\citenamefont {B\"ottcher}, \citenamefont {Schmidt}, \citenamefont {Wenzel}, \citenamefont {Hertkorn}, \citenamefont {Guo}, \citenamefont {Langen},\ and\ \citenamefont {Pfau}}]{Boettcher2019}%
  \BibitemOpen
  \bibfield  {author} {\bibinfo {author} {\bibfnamefont {F.}~\bibnamefont {B\"ottcher}}, \bibinfo {author} {\bibfnamefont {J.-N.}\ \bibnamefont {Schmidt}}, \bibinfo {author} {\bibfnamefont {M.}~\bibnamefont {Wenzel}}, \bibinfo {author} {\bibfnamefont {J.}~\bibnamefont {Hertkorn}}, \bibinfo {author} {\bibfnamefont {M.}~\bibnamefont {Guo}}, \bibinfo {author} {\bibfnamefont {T.}~\bibnamefont {Langen}},\ and\ \bibinfo {author} {\bibfnamefont {T.}~\bibnamefont {Pfau}},\ }\bibfield  {title} {\bibinfo {title} {Transient supersolid properties in an array of dipolar quantum droplets},\ }\href {https://doi.org/10.1103/PhysRevX.9.011051} {\bibfield  {journal} {\bibinfo  {journal} {Phys. Rev. X}\ }\textbf {\bibinfo {volume} {9}},\ \bibinfo {pages} {011051} (\bibinfo {year} {2019})}\BibitemShut {NoStop}%
\bibitem [{\citenamefont {Chomaz}\ \emph {et~al.}(2019)\citenamefont {Chomaz}, \citenamefont {Petter}, \citenamefont {Ilzh\"ofer}, \citenamefont {Natale}, \citenamefont {Trautmann}, \citenamefont {Politi}, \citenamefont {Durastante}, \citenamefont {van Bijnen}, \citenamefont {Patscheider}, \citenamefont {Sohmen}, \citenamefont {Mark},\ and\ \citenamefont {Ferlaino}}]{Chomaz2019}%
  \BibitemOpen
  \bibfield  {author} {\bibinfo {author} {\bibfnamefont {L.}~\bibnamefont {Chomaz}}, \bibinfo {author} {\bibfnamefont {D.}~\bibnamefont {Petter}}, \bibinfo {author} {\bibfnamefont {P.}~\bibnamefont {Ilzh\"ofer}}, \bibinfo {author} {\bibfnamefont {G.}~\bibnamefont {Natale}}, \bibinfo {author} {\bibfnamefont {A.}~\bibnamefont {Trautmann}}, \bibinfo {author} {\bibfnamefont {C.}~\bibnamefont {Politi}}, \bibinfo {author} {\bibfnamefont {G.}~\bibnamefont {Durastante}}, \bibinfo {author} {\bibfnamefont {R.~M.~W.}\ \bibnamefont {van Bijnen}}, \bibinfo {author} {\bibfnamefont {A.}~\bibnamefont {Patscheider}}, \bibinfo {author} {\bibfnamefont {M.}~\bibnamefont {Sohmen}}, \bibinfo {author} {\bibfnamefont {M.~J.}\ \bibnamefont {Mark}},\ and\ \bibinfo {author} {\bibfnamefont {F.}~\bibnamefont {Ferlaino}},\ }\bibfield  {title} {\bibinfo {title} {Long-lived and transient supersolid behaviors in dipolar quantum gases},\ }\href {https://doi.org/10.1103/PhysRevX.9.021012} {\bibfield  {journal} {\bibinfo  {journal} {Phys. Rev.
  X}\ }\textbf {\bibinfo {volume} {9}},\ \bibinfo {pages} {021012} (\bibinfo {year} {2019})}\BibitemShut {NoStop}%
\bibitem [{\citenamefont {Lin}\ \emph {et~al.}(2011{\natexlab{a}})\citenamefont {Lin}, \citenamefont {Jim{\'{e}}nez-Garc{\'{i}}a},\ and\ \citenamefont {Spielman}}]{Lin2011}%
  \BibitemOpen
  \bibfield  {author} {\bibinfo {author} {\bibfnamefont {Y.~J.}\ \bibnamefont {Lin}}, \bibinfo {author} {\bibfnamefont {K.}~\bibnamefont {Jim{\'{e}}nez-Garc{\'{i}}a}},\ and\ \bibinfo {author} {\bibfnamefont {I.~B.}\ \bibnamefont {Spielman}},\ }\bibfield  {title} {\bibinfo {title} {{Spin-orbit-coupled Bose-Einstein condensates}},\ }\href {https://doi.org/10.1038/nature09887} {\bibfield  {journal} {\bibinfo  {journal} {Nature}\ }\textbf {\bibinfo {volume} {471}},\ \bibinfo {pages} {83} (\bibinfo {year} {2011}{\natexlab{a}})}\BibitemShut {NoStop}%
\bibitem [{\citenamefont {Wang}\ \emph {et~al.}(2010)\citenamefont {Wang}, \citenamefont {Gao}, \citenamefont {Jian},\ and\ \citenamefont {Zhai}}]{Wang2010}%
  \BibitemOpen
  \bibfield  {author} {\bibinfo {author} {\bibfnamefont {C.}~\bibnamefont {Wang}}, \bibinfo {author} {\bibfnamefont {C.}~\bibnamefont {Gao}}, \bibinfo {author} {\bibfnamefont {C.-M.}\ \bibnamefont {Jian}},\ and\ \bibinfo {author} {\bibfnamefont {H.}~\bibnamefont {Zhai}},\ }\bibfield  {title} {\bibinfo {title} {Spin-orbit coupled spinor {B}ose-{E}instein condensates},\ }\href {https://doi.org/10.1103/PhysRevLett.105.160403} {\bibfield  {journal} {\bibinfo  {journal} {Phys. Rev. Lett.}\ }\textbf {\bibinfo {volume} {105}},\ \bibinfo {pages} {160403} (\bibinfo {year} {2010})}\BibitemShut {NoStop}%
\bibitem [{\citenamefont {Ho}\ and\ \citenamefont {Zhang}(2011)}]{Ho2011}%
  \BibitemOpen
  \bibfield  {author} {\bibinfo {author} {\bibfnamefont {T.-L.}\ \bibnamefont {Ho}}\ and\ \bibinfo {author} {\bibfnamefont {S.}~\bibnamefont {Zhang}},\ }\bibfield  {title} {\bibinfo {title} {Bose-{E}instein condensates with spin-orbit interaction},\ }\href {https://doi.org/10.1103/PhysRevLett.107.150403} {\bibfield  {journal} {\bibinfo  {journal} {Phys. Rev. Lett.}\ }\textbf {\bibinfo {volume} {107}},\ \bibinfo {pages} {150403} (\bibinfo {year} {2011})}\BibitemShut {NoStop}%
\bibitem [{\citenamefont {Li}\ \emph {et~al.}(2012{\natexlab{a}})\citenamefont {Li}, \citenamefont {Pitaevskii},\ and\ \citenamefont {Stringari}}]{Li2012}%
  \BibitemOpen
  \bibfield  {author} {\bibinfo {author} {\bibfnamefont {Y.}~\bibnamefont {Li}}, \bibinfo {author} {\bibfnamefont {L.~P.}\ \bibnamefont {Pitaevskii}},\ and\ \bibinfo {author} {\bibfnamefont {S.}~\bibnamefont {Stringari}},\ }\bibfield  {title} {\bibinfo {title} {{Quantum tricriticality and phase transitions in spin-orbit coupled Bose-Einstein condensates}},\ }\href {https://doi.org/10.1103/PhysRevLett.108.225301} {\bibfield  {journal} {\bibinfo  {journal} {Phys. Rev. Lett.}\ }\textbf {\bibinfo {volume} {108}},\ \bibinfo {pages} {225301} (\bibinfo {year} {2012}{\natexlab{a}})}\BibitemShut {NoStop}%
\bibitem [{\citenamefont {Martone}\ and\ \citenamefont {Stringari}(2021)}]{Martone2021}%
  \BibitemOpen
  \bibfield  {author} {\bibinfo {author} {\bibfnamefont {G.~I.}\ \bibnamefont {Martone}}\ and\ \bibinfo {author} {\bibfnamefont {S.}~\bibnamefont {Stringari}},\ }\bibfield  {title} {\bibinfo {title} {{Supersolid phase of a spin-orbit-coupled Bose-Einstein condensate: A perturbation approach}},\ }\href {https://doi.org/10.21468/SCIPOSTPHYS.11.5.092} {\bibfield  {journal} {\bibinfo  {journal} {SciPost Phys.}\ }\textbf {\bibinfo {volume} {11}},\ \bibinfo {pages} {92} (\bibinfo {year} {2021})}\BibitemShut {NoStop}%
\bibitem [{\citenamefont {Martone}(2015)}]{Martone2015}%
  \BibitemOpen
  \bibfield  {author} {\bibinfo {author} {\bibfnamefont {G.~I.}\ \bibnamefont {Martone}},\ }\bibfield  {title} {\bibinfo {title} {Visibility and stability of superstripes in a spin-orbit-coupled {B}ose-{E}instein condensate},\ }\href {https://doi.org/10.1140/epjst/e2015-02386-x} {\bibfield  {journal} {\bibinfo  {journal} {Eur. Phys. J.: Spec. Top.}\ }\textbf {\bibinfo {volume} {224}},\ \bibinfo {pages} {553} (\bibinfo {year} {2015})}\BibitemShut {NoStop}%
\bibitem [{\citenamefont {{Hofmann}}\ and\ \citenamefont {{Zwerger}}(2021)}]{Hofmann2021}%
  \BibitemOpen
  \bibfield  {author} {\bibinfo {author} {\bibfnamefont {J.}~\bibnamefont {{Hofmann}}}\ and\ \bibinfo {author} {\bibfnamefont {W.}~\bibnamefont {{Zwerger}}},\ }\bibfield  {title} {\bibinfo {title} {{Hydrodynamics of a superfluid smectic}},\ }\href {https://doi.org/10.1088/1742-5468/abe598} {\bibfield  {journal} {\bibinfo  {journal} {J. Stat. Mech.}\ }\textbf {\bibinfo {volume} {2021}},\ \bibinfo {eid} {033104} (\bibinfo {year} {2021})}\BibitemShut {NoStop}%
\bibitem [{\citenamefont {L{\'e}onard}\ \emph {et~al.}(2017)\citenamefont {L{\'e}onard}, \citenamefont {Morales}, \citenamefont {Zupancic}, \citenamefont {Donner},\ and\ \citenamefont {Esslinger}}]{LeonardScience2017}%
  \BibitemOpen
  \bibfield  {author} {\bibinfo {author} {\bibfnamefont {J.}~\bibnamefont {L{\'e}onard}}, \bibinfo {author} {\bibfnamefont {A.}~\bibnamefont {Morales}}, \bibinfo {author} {\bibfnamefont {P.}~\bibnamefont {Zupancic}}, \bibinfo {author} {\bibfnamefont {T.}~\bibnamefont {Donner}},\ and\ \bibinfo {author} {\bibfnamefont {T.}~\bibnamefont {Esslinger}},\ }\bibfield  {title} {\bibinfo {title} {Monitoring and manipulating {H}iggs and {G}oldstone modes in a supersolid quantum gas},\ }\href {https://doi.org/10.1126/science.aan2608} {\bibfield  {journal} {\bibinfo  {journal} {Science}\ }\textbf {\bibinfo {volume} {358}},\ \bibinfo {pages} {1415} (\bibinfo {year} {2017})}\BibitemShut {NoStop}%
\bibitem [{\citenamefont {{Tanzi}}\ \emph {et~al.}(2019)\citenamefont {{Tanzi}}, \citenamefont {{Roccuzzo}}, \citenamefont {{Lucioni}}, \citenamefont {{Fam{\`a}}}, \citenamefont {{Fioretti}}, \citenamefont {{Gabbanini}}, \citenamefont {{Modugno}}, \citenamefont {{Recati}},\ and\ \citenamefont {{Stringari}}}]{TanziNature2019}%
  \BibitemOpen
  \bibfield  {author} {\bibinfo {author} {\bibfnamefont {L.}~\bibnamefont {{Tanzi}}}, \bibinfo {author} {\bibfnamefont {S.~M.}\ \bibnamefont {{Roccuzzo}}}, \bibinfo {author} {\bibfnamefont {E.}~\bibnamefont {{Lucioni}}}, \bibinfo {author} {\bibfnamefont {F.}~\bibnamefont {{Fam{\`a}}}}, \bibinfo {author} {\bibfnamefont {A.}~\bibnamefont {{Fioretti}}}, \bibinfo {author} {\bibfnamefont {C.}~\bibnamefont {{Gabbanini}}}, \bibinfo {author} {\bibfnamefont {G.}~\bibnamefont {{Modugno}}}, \bibinfo {author} {\bibfnamefont {A.}~\bibnamefont {{Recati}}},\ and\ \bibinfo {author} {\bibfnamefont {S.}~\bibnamefont {{Stringari}}},\ }\bibfield  {title} {\bibinfo {title} {{Supersolid symmetry breaking from compressional oscillations in a dipolar quantum gas}},\ }\href {https://doi.org/10.1038/s41586-019-1568-6} {\bibfield  {journal} {\bibinfo  {journal} {Nature}\ }\textbf {\bibinfo {volume} {574}},\ \bibinfo {pages} {382} (\bibinfo {year} {2019})}\BibitemShut {NoStop}%
\bibitem [{\citenamefont {{Guo}}\ \emph {et~al.}(2019)\citenamefont {{Guo}}, \citenamefont {{B{\"o}ttcher}}, \citenamefont {{Hertkorn}}, \citenamefont {{Schmidt}}, \citenamefont {{Wenzel}}, \citenamefont {{B{\"u}chler}}, \citenamefont {{Langen}},\ and\ \citenamefont {{Pfau}}}]{Guo2019}%
  \BibitemOpen
  \bibfield  {author} {\bibinfo {author} {\bibfnamefont {M.}~\bibnamefont {{Guo}}}, \bibinfo {author} {\bibfnamefont {F.}~\bibnamefont {{B{\"o}ttcher}}}, \bibinfo {author} {\bibfnamefont {J.}~\bibnamefont {{Hertkorn}}}, \bibinfo {author} {\bibfnamefont {J.-N.}\ \bibnamefont {{Schmidt}}}, \bibinfo {author} {\bibfnamefont {M.}~\bibnamefont {{Wenzel}}}, \bibinfo {author} {\bibfnamefont {H.~P.}\ \bibnamefont {{B{\"u}chler}}}, \bibinfo {author} {\bibfnamefont {T.}~\bibnamefont {{Langen}}},\ and\ \bibinfo {author} {\bibfnamefont {T.}~\bibnamefont {{Pfau}}},\ }\bibfield  {title} {\bibinfo {title} {{The low-energy Goldstone mode in a trapped dipolar supersolid}},\ }\href {https://doi.org/10.1038/s41586-019-1569-5} {\bibfield  {journal} {\bibinfo  {journal} {Nature}\ }\textbf {\bibinfo {volume} {574}},\ \bibinfo {pages} {386} (\bibinfo {year} {2019})}\BibitemShut {NoStop}%
\bibitem [{\citenamefont {Natale}\ \emph {et~al.}(2019)\citenamefont {Natale}, \citenamefont {van Bijnen}, \citenamefont {Patscheider}, \citenamefont {Petter}, \citenamefont {Mark}, \citenamefont {Chomaz},\ and\ \citenamefont {Ferlaino}}]{Natale2019}%
  \BibitemOpen
  \bibfield  {author} {\bibinfo {author} {\bibfnamefont {G.}~\bibnamefont {Natale}}, \bibinfo {author} {\bibfnamefont {R.~M.~W.}\ \bibnamefont {van Bijnen}}, \bibinfo {author} {\bibfnamefont {A.}~\bibnamefont {Patscheider}}, \bibinfo {author} {\bibfnamefont {D.}~\bibnamefont {Petter}}, \bibinfo {author} {\bibfnamefont {M.~J.}\ \bibnamefont {Mark}}, \bibinfo {author} {\bibfnamefont {L.}~\bibnamefont {Chomaz}},\ and\ \bibinfo {author} {\bibfnamefont {F.}~\bibnamefont {Ferlaino}},\ }\bibfield  {title} {\bibinfo {title} {Excitation spectrum of a trapped dipolar supersolid and its experimental evidence},\ }\href {https://doi.org/10.1103/PhysRevLett.123.050402} {\bibfield  {journal} {\bibinfo  {journal} {Phys. Rev. Lett.}\ }\textbf {\bibinfo {volume} {123}},\ \bibinfo {pages} {050402} (\bibinfo {year} {2019})}\BibitemShut {NoStop}%
\bibitem [{\citenamefont {Tanzi}\ \emph {et~al.}(2021)\citenamefont {Tanzi}, \citenamefont {Maloberti}, \citenamefont {Biagioni}, \citenamefont {Fioretti}, \citenamefont {Gabbanini},\ and\ \citenamefont {Modugno}}]{Tanzi2021}%
  \BibitemOpen
  \bibfield  {author} {\bibinfo {author} {\bibfnamefont {L.}~\bibnamefont {Tanzi}}, \bibinfo {author} {\bibfnamefont {J.~G.}\ \bibnamefont {Maloberti}}, \bibinfo {author} {\bibfnamefont {G.}~\bibnamefont {Biagioni}}, \bibinfo {author} {\bibfnamefont {A.}~\bibnamefont {Fioretti}}, \bibinfo {author} {\bibfnamefont {C.}~\bibnamefont {Gabbanini}},\ and\ \bibinfo {author} {\bibfnamefont {G.}~\bibnamefont {Modugno}},\ }\bibfield  {title} {\bibinfo {title} {Evidence of superfluidity in a dipolar supersolid from nonclassical rotational inertia},\ }\href {https://doi.org/10.1126/science.aba4309} {\bibfield  {journal} {\bibinfo  {journal} {Science}\ }\textbf {\bibinfo {volume} {371}},\ \bibinfo {pages} {1162–1165} (\bibinfo {year} {2021})}\BibitemShut {NoStop}%
\bibitem [{\citenamefont {Norcia}\ \emph {et~al.}(2022)\citenamefont {Norcia}, \citenamefont {Poli}, \citenamefont {Politi}, \citenamefont {Klaus}, \citenamefont {Bland}, \citenamefont {Mark}, \citenamefont {Santos}, \citenamefont {Bisset},\ and\ \citenamefont {Ferlaino}}]{Norcia2022}%
  \BibitemOpen
  \bibfield  {author} {\bibinfo {author} {\bibfnamefont {M.~A.}\ \bibnamefont {Norcia}}, \bibinfo {author} {\bibfnamefont {E.}~\bibnamefont {Poli}}, \bibinfo {author} {\bibfnamefont {C.}~\bibnamefont {Politi}}, \bibinfo {author} {\bibfnamefont {L.}~\bibnamefont {Klaus}}, \bibinfo {author} {\bibfnamefont {T.}~\bibnamefont {Bland}}, \bibinfo {author} {\bibfnamefont {M.~J.}\ \bibnamefont {Mark}}, \bibinfo {author} {\bibfnamefont {L.}~\bibnamefont {Santos}}, \bibinfo {author} {\bibfnamefont {R.~N.}\ \bibnamefont {Bisset}},\ and\ \bibinfo {author} {\bibfnamefont {F.}~\bibnamefont {Ferlaino}},\ }\bibfield  {title} {\bibinfo {title} {Can angular oscillations probe superfluidity in dipolar supersolids?},\ }\href {https://doi.org/10.1103/PhysRevLett.129.040403} {\bibfield  {journal} {\bibinfo  {journal} {Phys. Rev. Lett.}\ }\textbf {\bibinfo {volume} {129}},\ \bibinfo {pages} {040403} (\bibinfo {year} {2022})}\BibitemShut {NoStop}%
\bibitem [{\citenamefont {Biagioni}\ \emph {et~al.}(2024)\citenamefont {Biagioni}, \citenamefont {Antolini}, \citenamefont {Donelli}, \citenamefont {Pezz{\`e}}, \citenamefont {Smerzi}, \citenamefont {Fattori}, \citenamefont {Fioretti}, \citenamefont {Gabbanini}, \citenamefont {Inguscio}, \citenamefont {Tanzi},\ and\ \citenamefont {Modugno}}]{Biagioni2024}%
  \BibitemOpen
  \bibfield  {author} {\bibinfo {author} {\bibfnamefont {G.}~\bibnamefont {Biagioni}}, \bibinfo {author} {\bibfnamefont {N.}~\bibnamefont {Antolini}}, \bibinfo {author} {\bibfnamefont {B.}~\bibnamefont {Donelli}}, \bibinfo {author} {\bibfnamefont {L.}~\bibnamefont {Pezz{\`e}}}, \bibinfo {author} {\bibfnamefont {A.}~\bibnamefont {Smerzi}}, \bibinfo {author} {\bibfnamefont {M.}~\bibnamefont {Fattori}}, \bibinfo {author} {\bibfnamefont {A.}~\bibnamefont {Fioretti}}, \bibinfo {author} {\bibfnamefont {C.}~\bibnamefont {Gabbanini}}, \bibinfo {author} {\bibfnamefont {M.}~\bibnamefont {Inguscio}}, \bibinfo {author} {\bibfnamefont {L.}~\bibnamefont {Tanzi}},\ and\ \bibinfo {author} {\bibfnamefont {G.}~\bibnamefont {Modugno}},\ }\bibfield  {title} {\bibinfo {title} {Measurement of the superfluid fraction of a supersolid by {J}osephson effect},\ }\href {https://doi.org/10.1038/s41586-024-07361-9} {\bibfield  {journal} {\bibinfo  {journal} {Nature}\ }\textbf {\bibinfo {volume} {629}},\ \bibinfo {pages} {773} (\bibinfo
  {year} {2024})}\BibitemShut {NoStop}%
\bibitem [{\citenamefont {Casotti}\ \emph {et~al.}(2024)\citenamefont {Casotti}, \citenamefont {Poli}, \citenamefont {Klaus}, \citenamefont {Litvinov}, \citenamefont {Ulm}, \citenamefont {Politi}, \citenamefont {Mark}, \citenamefont {Bland},\ and\ \citenamefont {Ferlaino}}]{Casotti2024}%
  \BibitemOpen
  \bibfield  {author} {\bibinfo {author} {\bibfnamefont {E.}~\bibnamefont {Casotti}}, \bibinfo {author} {\bibfnamefont {E.}~\bibnamefont {Poli}}, \bibinfo {author} {\bibfnamefont {L.}~\bibnamefont {Klaus}}, \bibinfo {author} {\bibfnamefont {A.}~\bibnamefont {Litvinov}}, \bibinfo {author} {\bibfnamefont {C.}~\bibnamefont {Ulm}}, \bibinfo {author} {\bibfnamefont {C.}~\bibnamefont {Politi}}, \bibinfo {author} {\bibfnamefont {M.~J.}\ \bibnamefont {Mark}}, \bibinfo {author} {\bibfnamefont {T.}~\bibnamefont {Bland}},\ and\ \bibinfo {author} {\bibfnamefont {F.}~\bibnamefont {Ferlaino}},\ }\bibfield  {title} {\bibinfo {title} {Observation of vortices in a dipolar supersolid},\ }\href {https://doi.org/10.1038/s41586-024-08149-7} {\bibfield  {journal} {\bibinfo  {journal} {Nature}\ }\textbf {\bibinfo {volume} {635}},\ \bibinfo {pages} {327} (\bibinfo {year} {2024})}\BibitemShut {NoStop}%
\bibitem [{\citenamefont {Putra}\ \emph {et~al.}(2020)\citenamefont {Putra}, \citenamefont {Salces-C{\'{a}}rcoba}, \citenamefont {Yue}, \citenamefont {Sugawa},\ and\ \citenamefont {Spielman}}]{Putra2020}%
  \BibitemOpen
  \bibfield  {author} {\bibinfo {author} {\bibfnamefont {A.}~\bibnamefont {Putra}}, \bibinfo {author} {\bibfnamefont {F.}~\bibnamefont {Salces-C{\'{a}}rcoba}}, \bibinfo {author} {\bibfnamefont {Y.}~\bibnamefont {Yue}}, \bibinfo {author} {\bibfnamefont {S.}~\bibnamefont {Sugawa}},\ and\ \bibinfo {author} {\bibfnamefont {I.~B.}\ \bibnamefont {Spielman}},\ }\bibfield  {title} {\bibinfo {title} {Spatial coherence of spin-orbit-coupled {B}ose gases},\ }\href {https://doi.org/10.1103/PhysRevLett.124.053605} {\bibfield  {journal} {\bibinfo  {journal} {Phys. Rev. Lett.}\ }\textbf {\bibinfo {volume} {124}},\ \bibinfo {pages} {053605} (\bibinfo {year} {2020})}\BibitemShut {NoStop}%
\bibitem [{\citenamefont {Geier}\ \emph {et~al.}(2023)\citenamefont {Geier}, \citenamefont {Martone}, \citenamefont {Hauke}, \citenamefont {Ketterle},\ and\ \citenamefont {Stringari}}]{Geier2023}%
  \BibitemOpen
  \bibfield  {author} {\bibinfo {author} {\bibfnamefont {K.~T.}\ \bibnamefont {Geier}}, \bibinfo {author} {\bibfnamefont {G.~I.}\ \bibnamefont {Martone}}, \bibinfo {author} {\bibfnamefont {P.}~\bibnamefont {Hauke}}, \bibinfo {author} {\bibfnamefont {W.}~\bibnamefont {Ketterle}},\ and\ \bibinfo {author} {\bibfnamefont {S.}~\bibnamefont {Stringari}},\ }\bibfield  {title} {\bibinfo {title} {Dynamics of stripe patterns in supersolid spin-orbit-coupled {B}ose gases},\ }\href {https://doi.org/10.1103/PhysRevLett.130.156001} {\bibfield  {journal} {\bibinfo  {journal} {Phys. Rev. Lett.}\ }\textbf {\bibinfo {volume} {130}},\ \bibinfo {pages} {156001} (\bibinfo {year} {2023})}\BibitemShut {NoStop}%
\bibitem [{\citenamefont {Tanzi}\ \emph {et~al.}(2018)\citenamefont {Tanzi}, \citenamefont {Cabrera}, \citenamefont {Sanz}, \citenamefont {Cheiney}, \citenamefont {Tomza},\ and\ \citenamefont {Tarruell}}]{Tanzi2018}%
  \BibitemOpen
  \bibfield  {author} {\bibinfo {author} {\bibfnamefont {L.}~\bibnamefont {Tanzi}}, \bibinfo {author} {\bibfnamefont {C.~R.}\ \bibnamefont {Cabrera}}, \bibinfo {author} {\bibfnamefont {J.}~\bibnamefont {Sanz}}, \bibinfo {author} {\bibfnamefont {P.}~\bibnamefont {Cheiney}}, \bibinfo {author} {\bibfnamefont {M.}~\bibnamefont {Tomza}},\ and\ \bibinfo {author} {\bibfnamefont {L.}~\bibnamefont {Tarruell}},\ }\bibfield  {title} {\bibinfo {title} {{Feshbach resonances in potassium Bose-Bose mixtures}},\ }\href {https://doi.org/10.1103/PhysRevA.98.062712} {\bibfield  {journal} {\bibinfo  {journal} {Phys. Rev. A}\ }\textbf {\bibinfo {volume} {98}},\ \bibinfo {pages} {062712} (\bibinfo {year} {2018})}\BibitemShut {NoStop}%
\bibitem [{\citenamefont {Higbie}\ and\ \citenamefont {Stamper-Kurn}(2004)}]{Higbie2004}%
  \BibitemOpen
  \bibfield  {author} {\bibinfo {author} {\bibfnamefont {J.}~\bibnamefont {Higbie}}\ and\ \bibinfo {author} {\bibfnamefont {D.~M.}\ \bibnamefont {Stamper-Kurn}},\ }\bibfield  {title} {\bibinfo {title} {Generating macroscopic-quantum-superposition states in momentum and internal-state space from {B}ose-{E}instein condensates with repulsive interactions},\ }\href {https://doi.org/10.1103/PhysRevA.69.053605} {\bibfield  {journal} {\bibinfo  {journal} {Phys. Rev. A}\ }\textbf {\bibinfo {volume} {69}},\ \bibinfo {pages} {053605} (\bibinfo {year} {2004})}\BibitemShut {NoStop}%
\bibitem [{met()}]{methods}%
  \BibitemOpen
  \bibinfo {note} {Materials and methods are available as supplementary material}\BibitemShut {NoStop}%
\bibitem [{\citenamefont {{Ji}}\ \emph {et~al.}(2014)\citenamefont {{Ji}}, \citenamefont {{Zhang}}, \citenamefont {{Zhang}}, \citenamefont {{Du}}, \citenamefont {{Zheng}}, \citenamefont {{Deng}}, \citenamefont {{Zhai}}, \citenamefont {{Chen}},\ and\ \citenamefont {{Pan}}}]{Ji2014}%
  \BibitemOpen
  \bibfield  {author} {\bibinfo {author} {\bibfnamefont {S.-C.}\ \bibnamefont {{Ji}}}, \bibinfo {author} {\bibfnamefont {J.-Y.}\ \bibnamefont {{Zhang}}}, \bibinfo {author} {\bibfnamefont {L.}~\bibnamefont {{Zhang}}}, \bibinfo {author} {\bibfnamefont {Z.-D.}\ \bibnamefont {{Du}}}, \bibinfo {author} {\bibfnamefont {W.}~\bibnamefont {{Zheng}}}, \bibinfo {author} {\bibfnamefont {Y.-J.}\ \bibnamefont {{Deng}}}, \bibinfo {author} {\bibfnamefont {H.}~\bibnamefont {{Zhai}}}, \bibinfo {author} {\bibfnamefont {S.}~\bibnamefont {{Chen}}},\ and\ \bibinfo {author} {\bibfnamefont {J.-W.}\ \bibnamefont {{Pan}}},\ }\bibfield  {title} {\bibinfo {title} {{Experimental determination of the finite-temperature phase diagram of a spin-orbit coupled Bose gas}},\ }\href {https://doi.org/10.1038/nphys2905} {\bibfield  {journal} {\bibinfo  {journal} {Nat. Phys.}\ }\textbf {\bibinfo {volume} {10}},\ \bibinfo {pages} {314} (\bibinfo {year} {2014})}\BibitemShut {NoStop}%
\bibitem [{\citenamefont {Bersano}\ \emph {et~al.}(2019)\citenamefont {Bersano}, \citenamefont {Hou}, \citenamefont {Mossman}, \citenamefont {Gokhroo}, \citenamefont {Luo}, \citenamefont {Sun}, \citenamefont {Zhang},\ and\ \citenamefont {Engels}}]{Bersano2019}%
  \BibitemOpen
  \bibfield  {author} {\bibinfo {author} {\bibfnamefont {T.~M.}\ \bibnamefont {Bersano}}, \bibinfo {author} {\bibfnamefont {J.}~\bibnamefont {Hou}}, \bibinfo {author} {\bibfnamefont {S.}~\bibnamefont {Mossman}}, \bibinfo {author} {\bibfnamefont {V.}~\bibnamefont {Gokhroo}}, \bibinfo {author} {\bibfnamefont {X.-W.}\ \bibnamefont {Luo}}, \bibinfo {author} {\bibfnamefont {K.}~\bibnamefont {Sun}}, \bibinfo {author} {\bibfnamefont {C.}~\bibnamefont {Zhang}},\ and\ \bibinfo {author} {\bibfnamefont {P.}~\bibnamefont {Engels}},\ }\bibfield  {title} {\bibinfo {title} {Experimental realization of a long-lived striped {B}ose-{E}instein condensate induced by momentum-space hopping},\ }\href {https://doi.org/10.1103/PhysRevA.99.051602} {\bibfield  {journal} {\bibinfo  {journal} {Phys. Rev. A}\ }\textbf {\bibinfo {volume} {99}},\ \bibinfo {pages} {051602} (\bibinfo {year} {2019})}\BibitemShut {NoStop}%
\bibitem [{\citenamefont {Chisholm}(2023)}]{Chisholm2023}%
  \BibitemOpen
  \bibfield  {author} {\bibinfo {author} {\bibfnamefont {C.~S.}\ \bibnamefont {Chisholm}},\ }\emph {\bibinfo {title} {{Raman dressed Bose-Einstein condensates with tunable interactions: topological gauge theories and supersolids}}},\ \href {https://doi.org/10.5821/dissertation-2117-393253} {Ph.D. thesis},\ \bibinfo  {school} {UPC, Institut de Ciències Fotòniques} (\bibinfo {year} {2023})\BibitemShut {NoStop}%
\bibitem [{\citenamefont {Menotti}\ and\ \citenamefont {Stringari}(2002)}]{Menotti2002}%
  \BibitemOpen
  \bibfield  {author} {\bibinfo {author} {\bibfnamefont {C.}~\bibnamefont {Menotti}}\ and\ \bibinfo {author} {\bibfnamefont {S.}~\bibnamefont {Stringari}},\ }\bibfield  {title} {\bibinfo {title} {Collective oscillations of a one-dimensional trapped {B}ose-{E}instein gas},\ }\href {https://doi.org/10.1103/PhysRevA.66.043610} {\bibfield  {journal} {\bibinfo  {journal} {Phys. Rev. A}\ }\textbf {\bibinfo {volume} {66}},\ \bibinfo {pages} {043610} (\bibinfo {year} {2002})}\BibitemShut {NoStop}%
\bibitem [{\citenamefont {Li}\ \emph {et~al.}(2012{\natexlab{b}})\citenamefont {Li}, \citenamefont {Martone},\ and\ \citenamefont {Stringari}}]{Li2012b}%
  \BibitemOpen
  \bibfield  {author} {\bibinfo {author} {\bibfnamefont {Y.}~\bibnamefont {Li}}, \bibinfo {author} {\bibfnamefont {G.~I.}\ \bibnamefont {Martone}},\ and\ \bibinfo {author} {\bibfnamefont {S.}~\bibnamefont {Stringari}},\ }\bibfield  {title} {\bibinfo {title} {Sum rules, dipole oscillation and spin polarizability of a spin-orbit coupled quantum gas},\ }\href {https://doi.org/10.1209/0295-5075/99/56008} {\bibfield  {journal} {\bibinfo  {journal} {EPL}\ }\textbf {\bibinfo {volume} {99}},\ \bibinfo {pages} {56008} (\bibinfo {year} {2012}{\natexlab{b}})}\BibitemShut {NoStop}%
\bibitem [{\citenamefont {Zhang}\ \emph {et~al.}(2012)\citenamefont {Zhang}, \citenamefont {Ji}, \citenamefont {Chen}, \citenamefont {Zhang}, \citenamefont {Du}, \citenamefont {Yan}, \citenamefont {Pan}, \citenamefont {Zhao}, \citenamefont {Deng}, \citenamefont {Zhai}, \citenamefont {Chen},\ and\ \citenamefont {Pan}}]{Zhang2012}%
  \BibitemOpen
  \bibfield  {author} {\bibinfo {author} {\bibfnamefont {J.-Y.}\ \bibnamefont {Zhang}}, \bibinfo {author} {\bibfnamefont {S.-C.}\ \bibnamefont {Ji}}, \bibinfo {author} {\bibfnamefont {Z.}~\bibnamefont {Chen}}, \bibinfo {author} {\bibfnamefont {L.}~\bibnamefont {Zhang}}, \bibinfo {author} {\bibfnamefont {Z.-D.}\ \bibnamefont {Du}}, \bibinfo {author} {\bibfnamefont {B.}~\bibnamefont {Yan}}, \bibinfo {author} {\bibfnamefont {G.-S.}\ \bibnamefont {Pan}}, \bibinfo {author} {\bibfnamefont {B.}~\bibnamefont {Zhao}}, \bibinfo {author} {\bibfnamefont {Y.-J.}\ \bibnamefont {Deng}}, \bibinfo {author} {\bibfnamefont {H.}~\bibnamefont {Zhai}}, \bibinfo {author} {\bibfnamefont {S.}~\bibnamefont {Chen}},\ and\ \bibinfo {author} {\bibfnamefont {J.-W.}\ \bibnamefont {Pan}},\ }\bibfield  {title} {\bibinfo {title} {Collective dipole oscillations of a spin-orbit coupled {B}ose-{E}instein condensate},\ }\href {https://doi.org/10.1103/PhysRevLett.109.115301} {\bibfield  {journal} {\bibinfo  {journal} {Phys. Rev. Lett.}\ }\textbf
  {\bibinfo {volume} {109}},\ \bibinfo {pages} {115301} (\bibinfo {year} {2012})}\BibitemShut {NoStop}%
\bibitem [{\citenamefont {Kr\"amer}\ \emph {et~al.}(2002)\citenamefont {Kr\"amer}, \citenamefont {Pitaevskii},\ and\ \citenamefont {Stringari}}]{Kraemer2002}%
  \BibitemOpen
  \bibfield  {author} {\bibinfo {author} {\bibfnamefont {M.}~\bibnamefont {Kr\"amer}}, \bibinfo {author} {\bibfnamefont {L.}~\bibnamefont {Pitaevskii}},\ and\ \bibinfo {author} {\bibfnamefont {S.}~\bibnamefont {Stringari}},\ }\bibfield  {title} {\bibinfo {title} {Macroscopic dynamics of a trapped {B}ose-{E}instein condensate in the presence of 1D and 2D optical lattices},\ }\href {https://doi.org/10.1103/PhysRevLett.88.180404} {\bibfield  {journal} {\bibinfo  {journal} {Phys. Rev. Lett.}\ }\textbf {\bibinfo {volume} {88}},\ \bibinfo {pages} {180404} (\bibinfo {year} {2002})}\BibitemShut {NoStop}%
\bibitem [{\citenamefont {Fort}\ \emph {et~al.}(2003)\citenamefont {Fort}, \citenamefont {Cataliotti}, \citenamefont {Fallani}, \citenamefont {Ferlaino}, \citenamefont {Maddaloni},\ and\ \citenamefont {Inguscio}}]{Fort2003}%
  \BibitemOpen
  \bibfield  {author} {\bibinfo {author} {\bibfnamefont {C.}~\bibnamefont {Fort}}, \bibinfo {author} {\bibfnamefont {F.~S.}\ \bibnamefont {Cataliotti}}, \bibinfo {author} {\bibfnamefont {L.}~\bibnamefont {Fallani}}, \bibinfo {author} {\bibfnamefont {F.}~\bibnamefont {Ferlaino}}, \bibinfo {author} {\bibfnamefont {P.}~\bibnamefont {Maddaloni}},\ and\ \bibinfo {author} {\bibfnamefont {M.}~\bibnamefont {Inguscio}},\ }\bibfield  {title} {\bibinfo {title} {Collective excitations of a trapped {B}ose-{E}instein condensate in the presence of a 1D optical lattice},\ }\href {https://doi.org/10.1103/PhysRevLett.90.140405} {\bibfield  {journal} {\bibinfo  {journal} {Phys. Rev. Lett.}\ }\textbf {\bibinfo {volume} {90}},\ \bibinfo {pages} {140405} (\bibinfo {year} {2003})}\BibitemShut {NoStop}%
\bibitem [{\citenamefont {{Guo}}\ \emph {et~al.}(2021)\citenamefont {{Guo}}, \citenamefont {{Kroeze}}, \citenamefont {{Marsh}}, \citenamefont {{Gopalakrishnan}}, \citenamefont {{Keeling}},\ and\ \citenamefont {{Lev}}}]{Guo2021}%
  \BibitemOpen
  \bibfield  {author} {\bibinfo {author} {\bibfnamefont {Y.}~\bibnamefont {{Guo}}}, \bibinfo {author} {\bibfnamefont {R.~M.}\ \bibnamefont {{Kroeze}}}, \bibinfo {author} {\bibfnamefont {B.~P.}\ \bibnamefont {{Marsh}}}, \bibinfo {author} {\bibfnamefont {S.}~\bibnamefont {{Gopalakrishnan}}}, \bibinfo {author} {\bibfnamefont {J.}~\bibnamefont {{Keeling}}},\ and\ \bibinfo {author} {\bibfnamefont {B.~L.}\ \bibnamefont {{Lev}}},\ }\bibfield  {title} {\bibinfo {title} {{An optical lattice with sound}},\ }\href {https://doi.org/10.1038/s41586-021-03945-x} {\bibfield  {journal} {\bibinfo  {journal} {Nature}\ }\textbf {\bibinfo {volume} {599}},\ \bibinfo {pages} {211} (\bibinfo {year} {2021})}\BibitemShut {NoStop}%
\bibitem [{\citenamefont {Li}\ \emph {et~al.}(2019)\citenamefont {Li}, \citenamefont {Qu}, \citenamefont {Niffenegger}, \citenamefont {Wang}, \citenamefont {He}, \citenamefont {Blasing}, \citenamefont {Olson}, \citenamefont {Greene}, \citenamefont {Lyanda-Geller}, \citenamefont {Zhou}, \citenamefont {Zhang},\ and\ \citenamefont {Chen}}]{Li2019}%
  \BibitemOpen
  \bibfield  {author} {\bibinfo {author} {\bibfnamefont {C.-H.}\ \bibnamefont {Li}}, \bibinfo {author} {\bibfnamefont {C.}~\bibnamefont {Qu}}, \bibinfo {author} {\bibfnamefont {R.~J.}\ \bibnamefont {Niffenegger}}, \bibinfo {author} {\bibfnamefont {S.-J.}\ \bibnamefont {Wang}}, \bibinfo {author} {\bibfnamefont {M.}~\bibnamefont {He}}, \bibinfo {author} {\bibfnamefont {D.~B.}\ \bibnamefont {Blasing}}, \bibinfo {author} {\bibfnamefont {A.~J.}\ \bibnamefont {Olson}}, \bibinfo {author} {\bibfnamefont {C.~H.}\ \bibnamefont {Greene}}, \bibinfo {author} {\bibfnamefont {Y.}~\bibnamefont {Lyanda-Geller}}, \bibinfo {author} {\bibfnamefont {Q.}~\bibnamefont {Zhou}}, \bibinfo {author} {\bibfnamefont {C.}~\bibnamefont {Zhang}},\ and\ \bibinfo {author} {\bibfnamefont {Y.~P.}\ \bibnamefont {Chen}},\ }\bibfield  {title} {\bibinfo {title} {Spin current generation and relaxation in a quenched spin-orbit-coupled Bose-Einstein condensate},\ }\href {https://doi.org/10.1038/s41467-018-08119-4} {\bibfield  {journal} {\bibinfo
  {journal} {Nat. Commun.}\ }\textbf {\bibinfo {volume} {10}},\ \bibinfo {pages} {375} (\bibinfo {year} {2019})}\BibitemShut {NoStop}%
\bibitem [{\citenamefont {Chen}\ \emph {et~al.}(2017)\citenamefont {Chen}, \citenamefont {Pu}, \citenamefont {Yu},\ and\ \citenamefont {Zhang}}]{Chen2017}%
  \BibitemOpen
  \bibfield  {author} {\bibinfo {author} {\bibfnamefont {L.}~\bibnamefont {Chen}}, \bibinfo {author} {\bibfnamefont {H.}~\bibnamefont {Pu}}, \bibinfo {author} {\bibfnamefont {Z.-Q.}\ \bibnamefont {Yu}},\ and\ \bibinfo {author} {\bibfnamefont {Y.}~\bibnamefont {Zhang}},\ }\bibfield  {title} {\bibinfo {title} {Collective excitation of a trapped {B}ose-{E}instein condensate with spin-orbit coupling},\ }\href {https://doi.org/10.1103/PhysRevA.95.033616} {\bibfield  {journal} {\bibinfo  {journal} {Phys. Rev. A}\ }\textbf {\bibinfo {volume} {95}},\ \bibinfo {pages} {033616} (\bibinfo {year} {2017})}\BibitemShut {NoStop}%
\bibitem [{\citenamefont {Geier}\ \emph {et~al.}(2021)\citenamefont {Geier}, \citenamefont {Martone}, \citenamefont {Hauke},\ and\ \citenamefont {Stringari}}]{Geier2021}%
  \BibitemOpen
  \bibfield  {author} {\bibinfo {author} {\bibfnamefont {K.~T.}\ \bibnamefont {Geier}}, \bibinfo {author} {\bibfnamefont {G.~I.}\ \bibnamefont {Martone}}, \bibinfo {author} {\bibfnamefont {P.}~\bibnamefont {Hauke}},\ and\ \bibinfo {author} {\bibfnamefont {S.}~\bibnamefont {Stringari}},\ }\bibfield  {title} {\bibinfo {title} {Exciting the {G}oldstone modes of a supersolid spin-orbit-coupled {B}ose gas},\ }\href {https://doi.org/10.1103/PhysRevLett.127.115301} {\bibfield  {journal} {\bibinfo  {journal} {Phys. Rev. Lett.}\ }\textbf {\bibinfo {volume} {127}},\ \bibinfo {pages} {115301} (\bibinfo {year} {2021})}\BibitemShut {NoStop}%
\bibitem [{\citenamefont {Lin}\ \emph {et~al.}(2011{\natexlab{b}})\citenamefont {Lin}, \citenamefont {Compton}, \citenamefont {Jim{\'e}nez-Garc{\'i}a}, \citenamefont {Phillips}, \citenamefont {Porto},\ and\ \citenamefont {Spielman}}]{Lin2011b}%
  \BibitemOpen
  \bibfield  {author} {\bibinfo {author} {\bibfnamefont {Y.-J.}\ \bibnamefont {Lin}}, \bibinfo {author} {\bibfnamefont {R.~L.}\ \bibnamefont {Compton}}, \bibinfo {author} {\bibfnamefont {K.}~\bibnamefont {Jim{\'e}nez-Garc{\'i}a}}, \bibinfo {author} {\bibfnamefont {W.~D.}\ \bibnamefont {Phillips}}, \bibinfo {author} {\bibfnamefont {J.~V.}\ \bibnamefont {Porto}},\ and\ \bibinfo {author} {\bibfnamefont {I.~B.}\ \bibnamefont {Spielman}},\ }\bibfield  {title} {\bibinfo {title} {A synthetic electric force acting on neutral atoms},\ }\href {https://doi.org/10.1038/nphys1954} {\bibfield  {journal} {\bibinfo  {journal} {Nat. Phys.}\ }\textbf {\bibinfo {volume} {7}},\ \bibinfo {pages} {531} (\bibinfo {year} {2011}{\natexlab{b}})}\BibitemShut {NoStop}%
\bibitem [{\citenamefont {Khamehchi}\ \emph {et~al.}(2017)\citenamefont {Khamehchi}, \citenamefont {Hossain}, \citenamefont {Mossman}, \citenamefont {Zhang}, \citenamefont {Busch}, \citenamefont {Forbes},\ and\ \citenamefont {Engels}}]{Khamehchi2017}%
  \BibitemOpen
  \bibfield  {author} {\bibinfo {author} {\bibfnamefont {M.~A.}\ \bibnamefont {Khamehchi}}, \bibinfo {author} {\bibfnamefont {K.}~\bibnamefont {Hossain}}, \bibinfo {author} {\bibfnamefont {M.~E.}\ \bibnamefont {Mossman}}, \bibinfo {author} {\bibfnamefont {Y.}~\bibnamefont {Zhang}}, \bibinfo {author} {\bibfnamefont {T.}~\bibnamefont {Busch}}, \bibinfo {author} {\bibfnamefont {M.~M.}\ \bibnamefont {Forbes}},\ and\ \bibinfo {author} {\bibfnamefont {P.}~\bibnamefont {Engels}},\ }\bibfield  {title} {\bibinfo {title} {Negative-mass hydrodynamics in a spin-orbit-coupled {B}ose-{E}instein condensate},\ }\href {https://doi.org/10.1103/PhysRevLett.118.155301} {\bibfield  {journal} {\bibinfo  {journal} {Phys. Rev. Lett.}\ }\textbf {\bibinfo {volume} {118}},\ \bibinfo {pages} {155301} (\bibinfo {year} {2017})}\BibitemShut {NoStop}%
\bibitem [{\citenamefont {Cavicchioli}\ \emph {et~al.}(2022)\citenamefont {Cavicchioli}, \citenamefont {Fort}, \citenamefont {Modugno}, \citenamefont {Minardi},\ and\ \citenamefont {Burchianti}}]{Cavicchioli2022}%
  \BibitemOpen
  \bibfield  {author} {\bibinfo {author} {\bibfnamefont {L.}~\bibnamefont {Cavicchioli}}, \bibinfo {author} {\bibfnamefont {C.}~\bibnamefont {Fort}}, \bibinfo {author} {\bibfnamefont {M.}~\bibnamefont {Modugno}}, \bibinfo {author} {\bibfnamefont {F.}~\bibnamefont {Minardi}},\ and\ \bibinfo {author} {\bibfnamefont {A.}~\bibnamefont {Burchianti}},\ }\bibfield  {title} {\bibinfo {title} {Dipole dynamics of an interacting bosonic mixture},\ }\href {https://doi.org/10.1103/PhysRevResearch.4.043068} {\bibfield  {journal} {\bibinfo  {journal} {Phys. Rev. Res.}\ }\textbf {\bibinfo {volume} {4}},\ \bibinfo {pages} {043068} (\bibinfo {year} {2022})}\BibitemShut {NoStop}%
\bibitem [{\citenamefont {Hertkorn}\ \emph {et~al.}(2019)\citenamefont {Hertkorn}, \citenamefont {B\"ottcher}, \citenamefont {Guo}, \citenamefont {Schmidt}, \citenamefont {Langen}, \citenamefont {B\"uchler},\ and\ \citenamefont {Pfau}}]{Hertkorn2019}%
  \BibitemOpen
  \bibfield  {author} {\bibinfo {author} {\bibfnamefont {J.}~\bibnamefont {Hertkorn}}, \bibinfo {author} {\bibfnamefont {F.}~\bibnamefont {B\"ottcher}}, \bibinfo {author} {\bibfnamefont {M.}~\bibnamefont {Guo}}, \bibinfo {author} {\bibfnamefont {J.~N.}\ \bibnamefont {Schmidt}}, \bibinfo {author} {\bibfnamefont {T.}~\bibnamefont {Langen}}, \bibinfo {author} {\bibfnamefont {H.~P.}\ \bibnamefont {B\"uchler}},\ and\ \bibinfo {author} {\bibfnamefont {T.}~\bibnamefont {Pfau}},\ }\bibfield  {title} {\bibinfo {title} {Fate of the amplitude mode in a trapped dipolar supersolid},\ }\href {https://doi.org/10.1103/PhysRevLett.123.193002} {\bibfield  {journal} {\bibinfo  {journal} {Phys. Rev. Lett.}\ }\textbf {\bibinfo {volume} {123}},\ \bibinfo {pages} {193002} (\bibinfo {year} {2019})}\BibitemShut {NoStop}%
\bibitem [{\citenamefont {Hertkorn}\ \emph {et~al.}(2024)\citenamefont {Hertkorn}, \citenamefont {St\"urmer}, \citenamefont {Mukherjee}, \citenamefont {Ng}, \citenamefont {Uerlings}, \citenamefont {Hellstern}, \citenamefont {Lavoine}, \citenamefont {Reimann}, \citenamefont {Pfau},\ and\ \citenamefont {Klemt}}]{Hertkorn2024}%
  \BibitemOpen
  \bibfield  {author} {\bibinfo {author} {\bibfnamefont {J.}~\bibnamefont {Hertkorn}}, \bibinfo {author} {\bibfnamefont {P.}~\bibnamefont {St\"urmer}}, \bibinfo {author} {\bibfnamefont {K.}~\bibnamefont {Mukherjee}}, \bibinfo {author} {\bibfnamefont {K.~S.~H.}\ \bibnamefont {Ng}}, \bibinfo {author} {\bibfnamefont {P.}~\bibnamefont {Uerlings}}, \bibinfo {author} {\bibfnamefont {F.}~\bibnamefont {Hellstern}}, \bibinfo {author} {\bibfnamefont {L.}~\bibnamefont {Lavoine}}, \bibinfo {author} {\bibfnamefont {S.~M.}\ \bibnamefont {Reimann}}, \bibinfo {author} {\bibfnamefont {T.}~\bibnamefont {Pfau}},\ and\ \bibinfo {author} {\bibfnamefont {R.}~\bibnamefont {Klemt}},\ }\bibfield  {title} {\bibinfo {title} {Decoupled sound and amplitude modes in trapped dipolar supersolids},\ }\href {https://doi.org/10.1103/PhysRevResearch.6.L042056} {\bibfield  {journal} {\bibinfo  {journal} {Phys. Rev. Res.}\ }\textbf {\bibinfo {volume} {6}},\ \bibinfo {pages} {L042056} (\bibinfo {year} {2024})}\BibitemShut {NoStop}%
\bibitem [{\citenamefont {Zhang}\ \emph {et~al.}(2016)\citenamefont {Zhang}, \citenamefont {Yu}, \citenamefont {Ng}, \citenamefont {Zhang}, \citenamefont {Pitaevskii},\ and\ \citenamefont {Stringari}}]{Zhang2016}%
  \BibitemOpen
  \bibfield  {author} {\bibinfo {author} {\bibfnamefont {Y.-C.}\ \bibnamefont {Zhang}}, \bibinfo {author} {\bibfnamefont {Z.-Q.}\ \bibnamefont {Yu}}, \bibinfo {author} {\bibfnamefont {T.~K.}\ \bibnamefont {Ng}}, \bibinfo {author} {\bibfnamefont {S.}~\bibnamefont {Zhang}}, \bibinfo {author} {\bibfnamefont {L.}~\bibnamefont {Pitaevskii}},\ and\ \bibinfo {author} {\bibfnamefont {S.}~\bibnamefont {Stringari}},\ }\bibfield  {title} {\bibinfo {title} {Superfluid density of a spin-orbit-coupled {B}ose gas},\ }\href {https://doi.org/10.1103/PhysRevA.94.033635} {\bibfield  {journal} {\bibinfo  {journal} {Phys. Rev. A}\ }\textbf {\bibinfo {volume} {94}},\ \bibinfo {pages} {033635} (\bibinfo {year} {2016})}\BibitemShut {NoStop}%
\bibitem [{\citenamefont {Cabrera}\ \emph {et~al.}(2018)\citenamefont {Cabrera}, \citenamefont {Tanzi}, \citenamefont {Sanz}, \citenamefont {Naylor}, \citenamefont {Thomas}, \citenamefont {Cheiney},\ and\ \citenamefont {Tarruell}}]{Cabrera2018}%
  \BibitemOpen
  \bibfield  {author} {\bibinfo {author} {\bibfnamefont {C.~R.}\ \bibnamefont {Cabrera}}, \bibinfo {author} {\bibfnamefont {L.}~\bibnamefont {Tanzi}}, \bibinfo {author} {\bibfnamefont {J.}~\bibnamefont {Sanz}}, \bibinfo {author} {\bibfnamefont {B.}~\bibnamefont {Naylor}}, \bibinfo {author} {\bibfnamefont {P.}~\bibnamefont {Thomas}}, \bibinfo {author} {\bibfnamefont {P.}~\bibnamefont {Cheiney}},\ and\ \bibinfo {author} {\bibfnamefont {L.}~\bibnamefont {Tarruell}},\ }\bibfield  {title} {\bibinfo {title} {{Quantum liquid droplets in a mixture of Bose-Einstein condensates}},\ }\href {https://doi.org/10.1126/science.aao5686} {\bibfield  {journal} {\bibinfo  {journal} {Science}\ }\textbf {\bibinfo {volume} {359}},\ \bibinfo {pages} {301} (\bibinfo {year} {2018})}\BibitemShut {NoStop}%
\bibitem [{\citenamefont {Semeghini}\ \emph {et~al.}(2018)\citenamefont {Semeghini}, \citenamefont {Ferioli}, \citenamefont {Masi}, \citenamefont {Mazzinghi}, \citenamefont {Wolswijk}, \citenamefont {Minardi}, \citenamefont {Modugno}, \citenamefont {Modugno}, \citenamefont {Inguscio},\ and\ \citenamefont {Fattori}}]{Semeghini2018}%
  \BibitemOpen
  \bibfield  {author} {\bibinfo {author} {\bibfnamefont {G.}~\bibnamefont {Semeghini}}, \bibinfo {author} {\bibfnamefont {G.}~\bibnamefont {Ferioli}}, \bibinfo {author} {\bibfnamefont {L.}~\bibnamefont {Masi}}, \bibinfo {author} {\bibfnamefont {C.}~\bibnamefont {Mazzinghi}}, \bibinfo {author} {\bibfnamefont {L.}~\bibnamefont {Wolswijk}}, \bibinfo {author} {\bibfnamefont {F.}~\bibnamefont {Minardi}}, \bibinfo {author} {\bibfnamefont {M.}~\bibnamefont {Modugno}}, \bibinfo {author} {\bibfnamefont {G.}~\bibnamefont {Modugno}}, \bibinfo {author} {\bibfnamefont {M.}~\bibnamefont {Inguscio}},\ and\ \bibinfo {author} {\bibfnamefont {M.}~\bibnamefont {Fattori}},\ }\bibfield  {title} {\bibinfo {title} {Self-bound quantum droplets of atomic mixtures in free space},\ }\href {https://doi.org/10.1103/PhysRevLett.120.235301} {\bibfield  {journal} {\bibinfo  {journal} {Phys. Rev. Lett.}\ }\textbf {\bibinfo {volume} {120}},\ \bibinfo {pages} {235301} (\bibinfo {year} {2018})}\BibitemShut {NoStop}%
\bibitem [{\citenamefont {Sachdeva}\ \emph {et~al.}(2020)\citenamefont {Sachdeva}, \citenamefont {Tengstrand},\ and\ \citenamefont {Reimann}}]{Sachdeva2020}%
  \BibitemOpen
  \bibfield  {author} {\bibinfo {author} {\bibfnamefont {R.}~\bibnamefont {Sachdeva}}, \bibinfo {author} {\bibfnamefont {M.~N.}\ \bibnamefont {Tengstrand}},\ and\ \bibinfo {author} {\bibfnamefont {S.~M.}\ \bibnamefont {Reimann}},\ }\bibfield  {title} {\bibinfo {title} {Self-bound supersolid stripe phase in binary {B}ose-{E}instein condensates},\ }\href {https://doi.org/10.1103/PhysRevA.102.043304} {\bibfield  {journal} {\bibinfo  {journal} {Phys. Rev. A}\ }\textbf {\bibinfo {volume} {102}},\ \bibinfo {pages} {043304} (\bibinfo {year} {2020})}\BibitemShut {NoStop}%
\bibitem [{\citenamefont {S{\'{a}}nchez-Baena}\ \emph {et~al.}(2020)\citenamefont {S{\'{a}}nchez-Baena}, \citenamefont {Boronat},\ and\ \citenamefont {Mazzanti}}]{Sanchez-Baena2020}%
  \BibitemOpen
  \bibfield  {author} {\bibinfo {author} {\bibfnamefont {J.}~\bibnamefont {S{\'{a}}nchez-Baena}}, \bibinfo {author} {\bibfnamefont {J.}~\bibnamefont {Boronat}},\ and\ \bibinfo {author} {\bibfnamefont {F.}~\bibnamefont {Mazzanti}},\ }\bibfield  {title} {\bibinfo {title} {{Supersolid striped droplets in a Raman spin-orbit-coupled system}},\ }\href {https://doi.org/10.1103/PhysRevA.102.053308} {\bibfield  {journal} {\bibinfo  {journal} {Phys. Rev. A}\ }\textbf {\bibinfo {volume} {102}},\ \bibinfo {pages} {053308} (\bibinfo {year} {2020})}\BibitemShut {NoStop}%
\bibitem [{\citenamefont {Chisholm}\ \emph {et~al.}(2025)\citenamefont {Chisholm}, \citenamefont {Hirthe}, \citenamefont {Makhalov}, \citenamefont {Ramos}, \citenamefont {Vatré}, \citenamefont {Cabedo}, \citenamefont {Celi},\ and\ \citenamefont {Tarruell}}]{Data}%
  \BibitemOpen
  \bibfield  {author} {\bibinfo {author} {\bibfnamefont {C.~S.}\ \bibnamefont {Chisholm}}, \bibinfo {author} {\bibfnamefont {S.}~\bibnamefont {Hirthe}}, \bibinfo {author} {\bibfnamefont {V.~B.}\ \bibnamefont {Makhalov}}, \bibinfo {author} {\bibfnamefont {R.}~\bibnamefont {Ramos}}, \bibinfo {author} {\bibfnamefont {R.}~\bibnamefont {Vatré}}, \bibinfo {author} {\bibfnamefont {J.}~\bibnamefont {Cabedo}}, \bibinfo {author} {\bibfnamefont {A.}~\bibnamefont {Celi}},\ and\ \bibinfo {author} {\bibfnamefont {L.}~\bibnamefont {Tarruell}},\ }\bibfield  {title} {\bibinfo {title} {Replication data for: Probing supersolidity through excitations in a spin-orbit-coupled Bose-Einstein condensate},\ }\bibfield  {journal} {\bibinfo  {journal} {CORA. Repositori de Dades de Recerca}\ }\href {https://doi.org/10.34810/data2791} {10.34810/data2791} (\bibinfo {year} {2025})\BibitemShut {NoStop}%

\end{thebibliography}

\begin{thebibliography}{25}

    \makeatletter
\providecommand \@ifxundefined [1]{%
 \@ifx{#1\undefined}
}%
\providecommand \@ifnum [1]{%
 \ifnum #1\expandafter \@firstoftwo
 \else \expandafter \@secondoftwo
 \fi
}%
\providecommand \@ifx [1]{%
 \ifx #1\expandafter \@firstoftwo
 \else \expandafter \@secondoftwo
 \fi
}%
\providecommand \natexlab [1]{#1}%
\providecommand \enquote  [1]{``#1''}%
\providecommand \bibnamefont  [1]{#1}%
\providecommand \bibfnamefont [1]{#1}%
\providecommand \citenamefont [1]{#1}%
\providecommand \@href[1]{\@@startlink{#1}\@@href}%
\providecommand \@@href[1]{\endgroup#1\@@endlink}%
\providecommand \@sanitize@url [0]{\catcode `\\12\catcode `\$12\catcode
  `\&12\catcode `\#12\catcode `\^12\catcode `\_12\catcode `\%12\relax}%
\providecommand \@@startlink[1]{}%
\providecommand \@@endlink[0]{}%
\providecommand \@url [1]{\endgroup\@href {#1}{\urlprefix }}%
\providecommand \urlprefix  [0]{URL }%
\providecommand \doibase [0]{https://doi.org/}%
\providecommand \selectlanguage [0]{\@gobble}%
\providecommand \bibinfo  [0]{\@secondoftwo}%
\providecommand \bibfield  [0]{\@secondoftwo}%
\providecommand \translation [1]{[#1]}%
\providecommand \BibitemOpen [0]{}%
\providecommand \bibitemStop [0]{}%
\providecommand \bibitemNoStop [0]{.\EOS\space}%
\providecommand \EOS [0]{\spacefactor3000\relax}%
\providecommand \BibitemShut  [1]{\csname bibitem#1\endcsname}%
\let\auto@bib@innerbib\@empty

\addtocounter{NAT@ctr}{53}

\bibitem [{\citenamefont {Fr\"{o}lian}\ \emph {et~al.}(2022)\citenamefont {Fr\"{o}lian}, \citenamefont {Chisholm}, \citenamefont {Neri}, \citenamefont {Cabrera}, \citenamefont {Ramos}, \citenamefont {Celi},\ and\ \citenamefont {Tarruell}}]{Froelian2022}%
  \BibitemOpen
  \bibfield  {author} {\bibinfo {author} {\bibfnamefont {A.}~\bibnamefont {Fr\"{o}lian}}, \bibinfo {author} {\bibfnamefont {C.~S.}\ \bibnamefont {Chisholm}}, \bibinfo {author} {\bibfnamefont {E.}~\bibnamefont {Neri}}, \bibinfo {author} {\bibfnamefont {C.~R.}\ \bibnamefont {Cabrera}}, \bibinfo {author} {\bibfnamefont {R.}~\bibnamefont {Ramos}}, \bibinfo {author} {\bibfnamefont {A.}~\bibnamefont {Celi}},\ and\ \bibinfo {author} {\bibfnamefont {L.}~\bibnamefont {Tarruell}},\ }\bibfield  {title} {\bibinfo {title} {{Realizing a 1D topological gauge theory in an optically dressed BEC}},\ }\href {https://doi.org/10.1038/s41586-022-04943-3} {\bibfield  {journal} {\bibinfo  {journal} {Nature}\ }\textbf {\bibinfo {volume} {608}},\ \bibinfo {pages} {293} (\bibinfo {year} {2022})}\BibitemShut {NoStop}%
\bibitem [{\citenamefont {Jiang}\ \emph {et~al.}(2013)\citenamefont {Jiang}, \citenamefont {Tang},\ and\ \citenamefont {Mitroy}}]{Jiang2013a}%
  \BibitemOpen
  \bibfield  {author} {\bibinfo {author} {\bibfnamefont {J.}~\bibnamefont {Jiang}}, \bibinfo {author} {\bibfnamefont {L.-Y.}\ \bibnamefont {Tang}},\ and\ \bibinfo {author} {\bibfnamefont {J.}~\bibnamefont {Mitroy}},\ }\bibfield  {title} {\bibinfo {title} {Tune-out wavelengths for potassium},\ }\href {https://doi.org/10.1103/PhysRevA.87.032518} {\bibfield  {journal} {\bibinfo  {journal} {Phys. Rev. A}\ }\textbf {\bibinfo {volume} {87}},\ \bibinfo {pages} {032518} (\bibinfo {year} {2013})}\BibitemShut {NoStop}%
\bibitem [{\citenamefont {Jiang}\ and\ \citenamefont {Mitroy}(2013)}]{Jiang2013b}%
  \BibitemOpen
  \bibfield  {author} {\bibinfo {author} {\bibfnamefont {J.}~\bibnamefont {Jiang}}\ and\ \bibinfo {author} {\bibfnamefont {J.}~\bibnamefont {Mitroy}},\ }\bibfield  {title} {\bibinfo {title} {Hyperfine effects on potassium tune-out wavelengths and polarizabilities},\ }\href {https://doi.org/10.1103/PhysRevA.88.032505} {\bibfield  {journal} {\bibinfo  {journal} {Phys. Rev. A}\ }\textbf {\bibinfo {volume} {88}},\ \bibinfo {pages} {032505} (\bibinfo {year} {2013})}\BibitemShut {NoStop}%
\bibitem [{\citenamefont {Shvarchuck}\ \emph {et~al.}(2002)\citenamefont {Shvarchuck}, \citenamefont {Buggle}, \citenamefont {Petrov}, \citenamefont {Dieckmann}, \citenamefont {Zielonkowski}, \citenamefont {Kemmann}, \citenamefont {Tiecke}, \citenamefont {von Klitzing}, \citenamefont {Shlyapnikov},\ and\ \citenamefont {Walraven}}]{Shvarchuck2002}%
  \BibitemOpen
  \bibfield  {author} {\bibinfo {author} {\bibfnamefont {I.}~\bibnamefont {Shvarchuck}}, \bibinfo {author} {\bibfnamefont {C.}~\bibnamefont {Buggle}}, \bibinfo {author} {\bibfnamefont {D.~S.}\ \bibnamefont {Petrov}}, \bibinfo {author} {\bibfnamefont {K.}~\bibnamefont {Dieckmann}}, \bibinfo {author} {\bibfnamefont {M.}~\bibnamefont {Zielonkowski}}, \bibinfo {author} {\bibfnamefont {M.}~\bibnamefont {Kemmann}}, \bibinfo {author} {\bibfnamefont {T.~G.}\ \bibnamefont {Tiecke}}, \bibinfo {author} {\bibfnamefont {W.}~\bibnamefont {von Klitzing}}, \bibinfo {author} {\bibfnamefont {G.~V.}\ \bibnamefont {Shlyapnikov}},\ and\ \bibinfo {author} {\bibfnamefont {J.~T.}\ \bibnamefont {Walraven}},\ }\bibfield  {title} {\bibinfo {title} {Bose-{E}instein condensation into nonequilibrium states studied by condensate focusing},\ }\href {https://doi.org/10.1103/PhysRevLett.89.270404} {\bibfield  {journal} {\bibinfo  {journal} {Phys. Rev. Lett.}\ }\textbf {\bibinfo {volume} {89}},\ \bibinfo {pages} {270404} (\bibinfo {year}
  {2002})}\BibitemShut {NoStop}%
\bibitem [{\citenamefont {Murthy}\ \emph {et~al.}(2014)\citenamefont {Murthy}, \citenamefont {Kedar}, \citenamefont {Lompe}, \citenamefont {Neidig}, \citenamefont {Ries}, \citenamefont {Wenz}, \citenamefont {Z{\"{u}}rn},\ and\ \citenamefont {Jochim}}]{Murthy2014}%
  \BibitemOpen
  \bibfield  {author} {\bibinfo {author} {\bibfnamefont {P.~A.}\ \bibnamefont {Murthy}}, \bibinfo {author} {\bibfnamefont {D.}~\bibnamefont {Kedar}}, \bibinfo {author} {\bibfnamefont {T.}~\bibnamefont {Lompe}}, \bibinfo {author} {\bibfnamefont {M.}~\bibnamefont {Neidig}}, \bibinfo {author} {\bibfnamefont {M.~G.}\ \bibnamefont {Ries}}, \bibinfo {author} {\bibfnamefont {A.~N.}\ \bibnamefont {Wenz}}, \bibinfo {author} {\bibfnamefont {G.}~\bibnamefont {Z{\"{u}}rn}},\ and\ \bibinfo {author} {\bibfnamefont {S.}~\bibnamefont {Jochim}},\ }\bibfield  {title} {\bibinfo {title} {{Matter-wave Fourier optics with a strongly interacting two-dimensional Fermi gas}},\ }\href {https://doi.org/10.1103/PhysRevA.90.043611} {\bibfield  {journal} {\bibinfo  {journal} {Phys. Rev. A}\ }\textbf {\bibinfo {volume} {90}},\ \bibinfo {pages} {043611} (\bibinfo {year} {2014})}\BibitemShut {NoStop}%
\bibitem [{\citenamefont {Asteria}\ \emph {et~al.}(2021)\citenamefont {Asteria}, \citenamefont {Zahn}, \citenamefont {Kosch}, \citenamefont {Sengstock},\ and\ \citenamefont {Weitenberg}}]{Asteria2021}%
  \BibitemOpen
  \bibfield  {author} {\bibinfo {author} {\bibfnamefont {L.}~\bibnamefont {Asteria}}, \bibinfo {author} {\bibfnamefont {H.~P.}\ \bibnamefont {Zahn}}, \bibinfo {author} {\bibfnamefont {M.~N.}\ \bibnamefont {Kosch}}, \bibinfo {author} {\bibfnamefont {K.}~\bibnamefont {Sengstock}},\ and\ \bibinfo {author} {\bibfnamefont {C.}~\bibnamefont {Weitenberg}},\ }\bibfield  {title} {\bibinfo {title} {{Quantum gas magnifier for sub-lattice-resolved imaging of 3D quantum systems}},\ }\href {https://doi.org/10.1038/s41586-021-04011-2} {\bibfield  {journal} {\bibinfo  {journal} {Nature}\ }\textbf {\bibinfo {volume} {599}},\ \bibinfo {pages} {571} (\bibinfo {year} {2021})}\BibitemShut {NoStop}%
\bibitem [{\citenamefont {Spielman}(2009)}]{Spielman2009}%
  \BibitemOpen
  \bibfield  {author} {\bibinfo {author} {\bibfnamefont {I.~B.}\ \bibnamefont {Spielman}},\ }\bibfield  {title} {\bibinfo {title} {Raman processes and effective gauge potentials},\ }\href {https://doi.org/10.1103/PhysRevA.79.063613} {\bibfield  {journal} {\bibinfo  {journal} {Phys. Rev. A}\ }\textbf {\bibinfo {volume} {79}},\ \bibinfo {pages} {063613} (\bibinfo {year} {2009})}\BibitemShut {NoStop}%
\bibitem [{\citenamefont {Williams}\ \emph {et~al.}(2012)\citenamefont {Williams}, \citenamefont {LeBlanc}, \citenamefont {Jiménez-García}, \citenamefont {Beeler}, \citenamefont {Perry}, \citenamefont {Phillips},\ and\ \citenamefont {Spielman}}]{Williams_science_2012}%
  \BibitemOpen
  \bibfield  {author} {\bibinfo {author} {\bibfnamefont {R.~A.}\ \bibnamefont {Williams}}, \bibinfo {author} {\bibfnamefont {L.~J.}\ \bibnamefont {LeBlanc}}, \bibinfo {author} {\bibfnamefont {K.}~\bibnamefont {Jiménez-García}}, \bibinfo {author} {\bibfnamefont {M.~C.}\ \bibnamefont {Beeler}}, \bibinfo {author} {\bibfnamefont {A.~R.}\ \bibnamefont {Perry}}, \bibinfo {author} {\bibfnamefont {W.~D.}\ \bibnamefont {Phillips}},\ and\ \bibinfo {author} {\bibfnamefont {I.~B.}\ \bibnamefont {Spielman}},\ }\bibfield  {title} {\bibinfo {title} {Synthetic partial waves in ultracold atomic collisions},\ }\href {https://doi.org/10.1126/science.1212652} {\bibfield  {journal} {\bibinfo  {journal} {Science}\ }\textbf {\bibinfo {volume} {335}},\ \bibinfo {pages} {314} (\bibinfo {year} {2012})}\BibitemShut {NoStop}%
\bibitem [{\citenamefont {Cabedo}\ \emph {et~al.}(2021)\citenamefont {Cabedo}, \citenamefont {Claramunt},\ and\ \citenamefont {Celi}}]{Cabedo2021a}%
  \BibitemOpen
  \bibfield  {author} {\bibinfo {author} {\bibfnamefont {J.}~\bibnamefont {Cabedo}}, \bibinfo {author} {\bibfnamefont {J.}~\bibnamefont {Claramunt}},\ and\ \bibinfo {author} {\bibfnamefont {A.}~\bibnamefont {Celi}},\ }\bibfield  {title} {\bibinfo {title} {{Dynamical preparation of stripe states in spin-orbit-coupled gases}},\ }\href {https://doi.org/10.1103/PhysRevA.104.L031305} {\bibfield  {journal} {\bibinfo  {journal} {Phys. Rev. A}\ }\textbf {\bibinfo {volume} {104}},\ \bibinfo {pages} {031305} (\bibinfo {year} {2021})}\BibitemShut {NoStop}%
\bibitem [{\citenamefont {Cabedo}\ and\ \citenamefont {Celi}(2021)}]{Cabedo2021b}%
  \BibitemOpen
  \bibfield  {author} {\bibinfo {author} {\bibfnamefont {J.}~\bibnamefont {Cabedo}}\ and\ \bibinfo {author} {\bibfnamefont {A.}~\bibnamefont {Celi}},\ }\bibfield  {title} {\bibinfo {title} {{Excited-state quantum phase transitions in spin-orbit-coupled Bose gases}},\ }\href {https://doi.org/10.1103/PhysRevResearch.3.043215} {\bibfield  {journal} {\bibinfo  {journal} {Phys. Rev. Res.}\ }\textbf {\bibinfo {volume} {3}},\ \bibinfo {pages} {043215} (\bibinfo {year} {2021})}\BibitemShut {NoStop}%
\bibitem [{\citenamefont {Chisholm}\ \emph {et~al.}(2022)\citenamefont {Chisholm}, \citenamefont {Fr\"olian}, \citenamefont {Neri}, \citenamefont {Ramos}, \citenamefont {Tarruell},\ and\ \citenamefont {Celi}}]{Chisholm2022}%
  \BibitemOpen
  \bibfield  {author} {\bibinfo {author} {\bibfnamefont {C.~S.}\ \bibnamefont {Chisholm}}, \bibinfo {author} {\bibfnamefont {A.}~\bibnamefont {Fr\"olian}}, \bibinfo {author} {\bibfnamefont {E.}~\bibnamefont {Neri}}, \bibinfo {author} {\bibfnamefont {R.}~\bibnamefont {Ramos}}, \bibinfo {author} {\bibfnamefont {L.}~\bibnamefont {Tarruell}},\ and\ \bibinfo {author} {\bibfnamefont {A.}~\bibnamefont {Celi}},\ }\bibfield  {title} {\bibinfo {title} {Encoding a one-dimensional topological gauge theory in a {R}aman-coupled {B}ose-{E}instein condensate},\ }\href {https://doi.org/10.1103/PhysRevResearch.4.043088} {\bibfield  {journal} {\bibinfo  {journal} {Phys. Rev. Res.}\ }\textbf {\bibinfo {volume} {4}},\ \bibinfo {pages} {043088} (\bibinfo {year} {2022})}\BibitemShut {NoStop}%
\bibitem [{\citenamefont {Fort}\ and\ \citenamefont {Modugno}(2022)}]{Fort2022}%
  \BibitemOpen
  \bibfield  {author} {\bibinfo {author} {\bibfnamefont {C.}~\bibnamefont {Fort}}\ and\ \bibinfo {author} {\bibfnamefont {M.}~\bibnamefont {Modugno}},\ }\bibfield  {title} {\bibinfo {title} {Dipole modes of a trapped bosonic mixture: Fate of the sum-rule approach},\ }\href {https://doi.org/10.1103/PhysRevA.106.043311} {\bibfield  {journal} {\bibinfo  {journal} {Phys. Rev. A}\ }\textbf {\bibinfo {volume} {106}},\ \bibinfo {pages} {043311} (\bibinfo {year} {2022})}\BibitemShut {NoStop}%

\end{thebibliography}
\end{document}